\documentclass[12pt]{iopart}

\usepackage{microtype}
\usepackage{graphicx}
\usepackage{subfigure}
\usepackage{booktabs}
\usepackage{makecell}
\usepackage{hyperref}
\usepackage{algorithmic}
\usepackage{algorithm}
\expandafter\let\csname equation*\endcsname\relax
\expandafter\let\csname endequation*\endcsname\relax
\usepackage{amsmath}
\usepackage{amssymb}
\usepackage{mathtools}
\usepackage{amsthm}
\usepackage{bm}
\usepackage[capitalize,noabbrev]{cleveref}
\theoremstyle{plain}

\theoremstyle{definition}

\theoremstyle{remark}

\usepackage[textsize=tiny]{todonotes}

\begin{document}

\title[Explorative Curriculum Learning for Strongly Correlated Electron Systems]{Explorative Curriculum Learning for Strongly Correlated Electron Systems}

\author{Kimihiro Yamazaki$^{1,2}$, Takuya Konishi$^{1,3}$, 
Yoshinobu Kawahara$^{1,3}$}

\address{$^1$ Graduate School of Information Science and Technology, The University of Osaka, 1-5 Yamadaoka, Suita, Osaka, Japan}
\address{$^2$ Fujitsu Limited, 4-1-1 Kamikodanaka, Nakahara-ku, Kawasaki-shi, Kanagawa, Japan}
\address{$^3$ Center for Advanced Intelligence Project, RIKEN, 1-4-1 Nihonbashi, Chuo-ku, Tokyo, Japan}
\ead{\{k-yamazaki, konishi, kawahara\}@ist.osaka-u.ac.jp}

\begin{abstract}

Recent advances in neural network quantum states (NQS) have enabled high-accuracy predictions for complex quantum many-body systems such as strongly correlated electron systems. However, the computational cost remains prohibitive, making exploration of the diverse parameters of interaction strengths and other physical parameters inefficient. While transfer learning has been proposed to mitigate this challenge, achieving generalization to large-scale systems and diverse parameter regimes remains difficult. To address this limitation, we propose a novel curriculum learning framework based on transfer learning for NQS. This facilitates efficient and stable exploration across a vast parameter space of quantum many-body systems. In addition, by interpreting NQS transfer learning through a perturbative lens, we demonstrate how prior physical knowledge can be flexibly incorporated into the curriculum learning process. We also propose Pairing-Net, an architecture to practically implement this strategy for strongly correlated electron systems, and empirically verify its effectiveness. Our results show an approximately 200-fold speedup in computation and a marked improvement in optimization stability compared to conventional methods.\\


\end{abstract}

\maketitle

\section{Introduction}
\label{introduction}

Obtaining quantum states of various quantum many-body systems is one of the most important problems in physics. A quantum many-body system consists of multiple interacting particles, and its properties are governed by the quantum state. Among these systems, strongly correlated electron systems are known to be particularly difficult to calculate those quantum states where many-body effects (i.e., electron correlation) play a crucial role~\cite{imada1998rev}. Such a strong correlation gives rise to diverse phenomena, e.g., magnetism~\cite{hirsch1985two} and high-temperature superconductivity~\cite{anderson1997theory}. Although the Hubbard model~\cite{hubbard1963electron} is widely studied as a prototypical model of strongly correlated electron systems, capturing its full complexity remains a grand challenge.

To tackle quantum many-body problems, methods utilizing neural network quantum states (NQS)~\cite{carleo2017solving} have recently garnered significant attention. NQS employ highly expressive neural networks to represent quantum states, achieving high computational accuracy even for strongly correlated electron systems~\cite{nomura2017restricted}. However, NQS often suffer from high computational costs and potential instability.
In addition, the NQS of quantum many-body systems can be viewed as a function of various parameters such as the number of electrons and the strength of interactions, with different regimes of quantum phases. Therefore, exploration of quantum states across these regions is crucial but computationally expensive. To alleviate these issues, transfer learning approaches have been proposed~\cite{10.5555/3618408.3618840,scherbela2023variational,zhu2023hubbardnet,scherbela2024towards,kim2024neural,gao2024neural,rende2024fine}; however, their effectiveness remains limited, especially for strongly correlated electron systems.

In this work, we aim to establish a machine learning method that significantly reduces the computational burden of exhaustively exploring NQS under diverse parameter regimes (e.g., varying electron numbers and interaction strengths). To this end, we propose a novel curriculum learning framework for the exhaustive exploration of multiple physical parameters. Conventional approaches suffer from prohibitive inefficiencies in such cases. In contrast, our method leverages the strategic iterative
application of transfer learning for NQS, enabling efficient and stable exploration across multiple physical parameters. Also, we provide a physical interpretation of this transfer learning process through the lens of perturbation theory: A physics methodology that incrementally renormalizes information from a simpler model to a more complex one. This perspective offers a clear physical interpretation of zero-shot learning and fine-tuning within the context of transfer learning. Building these insights, we propose Pairing-Net, a transferable NQS architecture that demonstrates how our curriculum learning framework can be practically applied to strongly correlated electron systems. Through several numerical experiments, we have demonstrated that our method can efficiently and stably explore a broad range of electronic correlation regimes in strongly correlated electron systems. Our method is applicable to larger systems compared to the previous study~\cite{zhu2023hubbardnet}, achieving over 200-fold speedup in computation with high computational stability. Furthermore, by conducting several performance evaluations of transfer learning, we confirmed that our method can acquire the high generalization performance in the context of electronic correlations. These results suggest that our approach represents a promising step toward applying transfer learning in strongly correlated electron systems, particularly within the framework of the Hubbard model.

The remainder of this paper is organized as follows: In Section~\ref{2}, we provide background on the quantum many-body problem and introduce NQS. In Section~\ref{3}, we discuss the limitations of previous works employing transfer learning to improve the computational efficiency of NQS. In Section~\ref{4}, we detail our curriculum learning method, designed to address these limitations. In Section~\ref{5}, we provide a physical interpretation of our method by examining the connection between transfer learning and perturbation theory. In Section~\ref{6}, we introduce Pairing-Net, an architecture for practically implementing our curriculum learning in strongly correlated electron systems. Numerical experiments validating the effectiveness of our method are shown in Section~\ref{7}, and finally we conclude the study in Section~\ref{8}.
Appendices include an overview of perturbation theory, and the details and additional results of the numerical experiments.

\section{Background}
\label{2}

\subsection{Quantum Many-Body Problem and Strongly Correlated Electron Systems}
\label{2-1}

The quantum many-body problem, meaning here the problem of obtaining quantum states in a quantum many-body system, is essentially reduced to solving the following eigenvalue problem for a Hamiltonian operator $\hat{\mathcal{H}}: \mathbb{H}\to\mathbb{H}$:
\begin{equation}
\hat{\mathcal{H}}|\Psi\rangle = E|\Psi\rangle,
\label{eq:eigen}
\end{equation}
where $\mathbb{H}$ is a Hilbert space, $|\Psi\rangle \in \mathbb{H}$ is an eigenvector (quantum state) corresponding to the Hamiltonian operator and $E \in \mathbb{R}$ is the eigen-energy corresponding to the eigenvector $|\Psi\rangle$. Eq.~(\ref{eq:eigen}) implies that the quantum state $|\Psi\rangle$ of the system is essentially determined by the Hamiltonian. 

The Hamiltonian operator is frequently approximated by employing a suitable finite basis set (e.g., a finite set of Fock states $\{|x\rangle\}$), which is known as an effective Hamiltonian. The total number of bases, denoted by $N_{B}$, corresponds to the dimension of the effective Hamiltonian, i.e., $\hat{H} \in \mathbb{C}^{N_B \times N_B}$. The restriction of the system to a finite-dimensional subspace corresponding to a specific energy range renders it numerically tractable, thereby reducing computational costs. Furthermore, the effective Hamiltonian can often be described as a function of physical parameters $\bm{\lambda}$ characterizing the system's properties, denoted as $\hat{H}(\bm{\lambda})$. The dimension of $\bm{\lambda}$, e.g., interaction strengths and the number of particles in the system, varies depending on the choice of an effective Hamiltonian.

The Hubbard model is a representative prototypical model to represent strongly correlated electron systems, and has been widely studied for almost half a century in physics~\cite{hubbard1963electron}, whose effective Hamiltonian is given by:\vspace*{-2.5mm}
\begin{equation}
\hat{H}(\bm{\lambda}) = -t \sum_{\langle i,j \rangle} \sum_{\sigma} \left( \hat{c}_{i\sigma}^{\dagger} \hat{c}_{j\sigma} + \text{H.c.}\right) + U \sum_{i=1}^N \hat{n}_{i\uparrow} \hat{n}_{i\downarrow},\vspace*{-1.5mm}
\label{eq:hubbard}
\end{equation}
where $\hat{c}_{i\sigma}^{\dagger}$ ($\hat{c}_{i\sigma}$) is the creation (annihilation) operator for an electron with spin $\sigma \in \{\uparrow, \downarrow\}$ at the spatial coordinate $\bm{r}_i \in \mathbb{R}^3$ of the $i$-th lattice site, and $\hat{n}_{i\sigma} \equiv \hat{c}_{i\sigma}^{\dagger} \hat{c}_{i\sigma}$ is the electron number operator. H.c. denotes the Hermitian conjugate of the preceding term, i.e., $\hat{c}_{j\sigma}^{\dagger} \hat{c}_{i\sigma} $. $t \in \mathbb{R}$ is the hopping integral (transition probability) of electrons between nearest-neighbor lattice sites, and $U \in \mathbb{R}$ is the on-site Coulomb repulsion interaction between electrons on the same lattice site. 
$N \in \mathbb{N}$ is the number of electrons, and the system size is determined by the number of lattice sites, $M \in \mathbb{N}$.

The difficulty of solving the quantum many-body problem for the Hubbard model can be understood through the eigenvalue problem~\eqref{eq:eigen}. 
For the Hubbard model of electron systems, $N_B = (2M)!/(2M-N)!N!$. In the case of a large-scale ($M \gg 1$) and near half-filling, where the electron density $\tilde{N} \equiv N/M$ equals 1, $N_B$ increases exponentially, resulting in a considerably high-dimensional optimization problem. Furthermore, both electron itinerancy and localization compete near half-filling as the ratio of Coulomb interaction to electron hopping, $\tilde{U} \equiv U/t$, changes, leading to complex electron correlations. This means that the physical parameters controlling the electron correlation in the model can be written as $\bm{\lambda} \equiv (\tilde{N}, \tilde{U})\in\mathbb{R}^{2}$. The physical interpretation of these parameters can be found in~\ref{Appendix-Hubbard}. 

In experimental investigations, the analysis of variations in $\tilde{U}$ or $\tilde{N}$ corresponds, for example, to examining variations in the quantum state resulting from crystal structure deformations~\cite{PhysRevB.99.195141,ijms24021509} or carrier doping~\cite{iimura2012two,PhysRevLett.113.027002}, respectively. Therefore, performing such analyses with high numerical accuracy and efficiency has emerged as a significant practical challenge.

\subsection{Variational Monte Carlo Method and Neural Network Quantum States}
\label{2-2}

The variational Monte Carlo (VMC) method~\cite{mcmillan1965ground} is often applied to find 
the quantum states for 
the lowest eigenvalue of Eq.~(\ref{eq:eigen}), which is called the ground state, for effective Hamiltonians. This method offers advantages over other methods in terms of applicability to large-scale systems and the fact that it is not directly hampered by the fermionic sign problem that plagues auxiliary-field quantum Monte Carlo (AF-QMC)~\cite{qin2016benchmark}. In the VMC method, a quantum state $|\Psi_{\theta}\rangle$ is parameterized by variational parameters $\theta$. The VMC method then determines the variational parameters $\theta$ that minimize the energy expectation value $E_{\theta}=\langle \Psi_\theta | \hat{H}(\bm{\lambda}) | \Psi_\theta \rangle/\langle \Psi_\theta | \Psi_\theta \rangle$, based on the Rayleigh-Ritz variational principle:
\begin{equation}
E_0 \le E_{\theta}=\frac{\sum_{x}|\Psi_{\theta}(x)|^{2}E^{(\mathrm{loc})}_{\theta}(x)}{\sum_{x}|\Psi_{\theta}(x)|^{2}},
\label{eq:variational}
\end{equation}
where $E_{0} \in \mathbb{R}$ denotes the minimum eigenvalue of the Hamiltonian, corresponding to the ground state energy. $\Psi_{\theta}(x)$ represents the wave function, defined as the coefficient $\Psi_{\theta}(x) \equiv \langle x | \Psi_{\theta} \rangle$ obtained by expanding $|\Psi_{\theta}\rangle$ in the basis set $\{|x\rangle\}$ as $|\Psi_{\theta}\rangle = \sum_{x} \Psi_{\theta}(x) |x\rangle$. Then, the squared wave function, $|\Psi_{\theta}(x)|^{2}$, provides a value proportional to the probability density corresponding to the basis $|x\rangle$. In this case, $E_\theta$ is calculated as the expectation value of the local energy $E^{(\mathrm{loc})}_{\theta}(x)\equiv \sum_{x'}\langle x|\hat{H}(\bm{\lambda})|x'\rangle\Psi_{\theta}(x')/\Psi_{\theta}(x)$, using $|\Psi_{\theta}(x)|^{2}$ as the probability density.

Eq.~(\ref{eq:variational}) can be viewed, in the context of machine learning, as an optimization problem for the energy expectation value parameterized by the variational parameters $\theta$. In the VMC method, $E_{\theta}$ is minimized using stochastic reconfiguration (natural gradient) method~\cite{amari1998natural,sorella2002superconductivity}. However, for high-complexity parameters, $N_B$ increases exponentially, making it difficult to compute the full summation over the bases in the energy expectation value. Therefore, $E_{\theta}=\mathbb{E}_{x\sim|\Psi_{\theta}(x)|^2}[E^{(\mathrm{loc})}_{\theta}(x)]$ is approximated using the Metropolis-Hastings algorithm~\cite{hastings1970monte} based on Markov chain Monte Carlo (MCMC):\vspace*{-1.5mm}
\begin{equation}
E_{\theta} \approx \frac{1}{N_{\mathrm{MC}}}\sum_{s=1}^{N_{\mathrm{MC}}}E^{(\mathrm{loc})}_{\theta}(x_{s}),\vspace*{-1.5mm}
\label{eq:mcmc}
\end{equation}
where $N_{\mathrm{MC}}$ represents the number of Monte Carlo samples. $x_{s}$ denotes each sample, which is often sampled from the real-space configuration of the particles. Thus, the $E_{\theta}$ can be efficiently calculated even with a vast number of bases. However, conventional VMC methods suffer from a critical drawback: The functional form of the wave function $|\Psi_{\theta}\rangle$ must be artificially designed and the validity of this design significantly impacts the results.

Neural network quantum states (NQS)~\cite{carleo2017solving} represent a groundbreaking approach that enables more expressive and flexible optimization by representing the functional form of $|\Psi_{\theta}\rangle$ with a neural network. However, the VMC method using NQS involves an optimization problem with a vast number of variational parameters, leading to high computational costs and inherent instability. Furthermore, conventional VMC-based approaches necessitate independent execution of computationally expensive optimization calculations for each physical parameter $\bm{\lambda}$. Consequently, as the number of physical parameters under consideration increases, the computational burden scales linearly, resulting in a highly inefficient exploration. These problems are particularly pronounced for complex quantum many-body systems such as strongly correlated electron systems.

\section{Related Works}
\label{3}

Conventional VMC-based NQS methods have been hampered by high computational costs and numerical instability. While recent works have demonstrated the potential of transfer learning to mitigate these issues across various quantum many-body systems~\cite{10.5555/3618408.3618840,scherbela2023variational,scherbela2024towards,kim2024neural,gao2024neural}, the effectiveness of transfer learning in this context remains limited.

Moreover, transfer learning methods have recently shown promise for efficient exploration of diverse physical properties across vast parameter spaces in quantum many-body systems~\cite{zhu2023hubbardnet,rende2024fine}. However, existing methods based on multi-task learning loss functions~\cite{zhu2023hubbardnet} exhibit limitations in generalization performance and large-scale systems. Recent work has also demonstrated the effectiveness of fine-tuning for exploring diverse quantum phases~\cite{rende2024fine}; however, identifying effective fine-tuning conditions necessary for exploration remains an open challenge. Furthermore, applications to strongly correlated electron systems necessitate satisfying the antisymmetry requirement of the wave function under fermionic particle exchange, demanding further innovations in transferable NQS architectures.

Finally, we remark on the relationship between existing fine-tuning methods~\cite{10.5555/3618408.3618840,scherbela2023variational,scherbela2024towards,kim2024neural,gao2024neural,rende2024fine} and the proposed method. Our method based on curriculum learning aims for efficient and stable exploration on the vast physical parameter space $\bm{\lambda}$ by iteratively fine-tuning in a ``proper" order. However, obtaining the appropriate order of fine-tuning is not straightforward in the existing methods. To address this, we demonstrate that the transfer learning can be interpreted as perturbation theory in physics, which provides a guideline for determining the appropriate fine-tuning order. This enables the establishment of an appropriate order in our curriculum learning.

\section{Curriculum Learning for Quantum Many-Body Problem}
\label{4}

Here we propose a novel curriculum learning framework aimed at improving the efficiency and stability of exploration of NQS across diverse parameters for quantum many-body systems.

\subsection{Transfer Learning with NQS}
\label{4-1}

We formalize the transfer learning of NQS, a crucial component of our proposed method. Conventional approaches with VMC necessitate independent training of NQS for each of the numerous physical parameters. Consequently, the exploration of these numerous parameters is computationally prohibitive. Our approach overcomes this limitation by employing NQS architectures capable of transfer learning, specifically fine-tuning pre-trained NQS across two different physical parameters $\bm{\lambda}, \bm{\lambda}'$.
The NQS architecture can be any architecture that enables knowledge transfer between arbitrary distinct physical parameters; Section~\ref{6} introduces Pairing-Net designed for strongly correlated electron systems. In the following, we introduce transfer learning (pre-training and fine-tuning) utilizing these transferable NQS. 

The task and transfer learning are defined as follows:\vspace*{-3mm}
\paragraph{Task:}
Task $\mathcal{T}(\bm{\lambda})$ is defined as the problem of obtaining the ground-state NQS $|\Psi_{\theta(\bm{\lambda})}\rangle$ by solving the optimization problem~(\ref{eq:variational}), where the loss function is the energy expectation value from Eq.~(\ref{eq:mcmc}) for the Hamiltonian $\hat{H}(\bm{\lambda})$ with physical parameter $\bm{\lambda}$. This is referred to as VMC calculation.
\vspace*{-3mm}
\paragraph{Pre-Training and Fine-Tuning:}
Let the pre-training task be $\mathcal{T}(\bm{\lambda})$. Then, the fine-tuning $\mathcal{F}(\bm{\lambda}\rightarrow\bm{\lambda}')$ to the target task $\mathcal{T}(\bm{\lambda}')$ can be formulated as follows. First, we define the pre-trained NQS as the initial state:
\begin{equation}
|\Psi_{\theta^{(0)}(\bm{\lambda'})}\rangle\equiv|\Psi_{\theta(\bm{\lambda})}\rangle.
\end{equation}
Fine-tuning is then defined as the operation of getting the ground-state NQS $|\Psi_{\theta(\bm{\lambda}')}\rangle$ through VMC calculation for the target Hamiltonian $\hat{H}(\bm{\lambda}')$, using this initial state.

\subsection{Curriculum Learning Algorithm}
\label{4-2}

\begin{algorithm}[tb]
    \caption{Explorative Curriculum Learning}
    \label{alg:algorithm}
    \textbf{Input}: Hamiltonians $\{\hat{H}_{k}\}_{k=1}^{K}$\\
    \textbf{Parameter}: Variational parameters of NQS $\{\theta_{k}\}_{k=1}^{K}$\\
    \textbf{Output}: Trained NQS $\{|\Psi_{\theta_{k}}\rangle\}_{k=1}^{K}$
    
    \begin{algorithmic}[1] 
    
        \FOR {$k=1,\cdots,K$}
        \IF {$k=1$}
        \STATE \# \textbf{Pre-Training}: $\mathcal{T}(\bm{\lambda}_{1})$
        \STATE Compute $|\Psi_{\theta_{1}}\rangle$ by VMC calculation with $\hat{H}_{1}$
        \ELSE
        \STATE \# \textbf{$k$-th Fine-Tuning}:  $\mathcal{F}(\bm{\lambda}_{k-1}\rightarrow\bm{\lambda}_{k})$
        \STATE Let $|\Psi_{\theta^{(0)}_{k}} \rangle=|\Psi_{\theta_{k-1}}\rangle$
        \STATE Compute $|\Psi_{\theta_{k}}\rangle$ by VMC calculation with $\hat{H}_{k}$
        \ENDIF
        \ENDFOR

    \end{algorithmic}
    \label{alg:ef-alg}
\end{algorithm}

Here we propose a novel curriculum learning approach that enables efficient and stable exhaustive exploration through iterative transfer learning. Suppose that we are first given a set of $K$ physical parameters in the parameter space of $\bm{\lambda}$ of our interest based on physical prior knowledge. We next determine an order of these parameters and obtain an ordered set of the parameters $\{\bm{\lambda}_k\}_{k=1}^{K}$. Then, our curriculum learning approach iteratively performs transfer learning for $\bm{\lambda}_1, \bm{\lambda}_2, \cdots, \bm{\lambda}_K$ in this order. For instance, in real materials, task sets—which represent a challenging domain in strongly correlated electron systems—may encompass monotonic variations of $\tilde{U}$ that result from crystal structure deformations induced by pressure, among other factors~\cite{PhysRevB.99.195141,ijms24021509}. Our method aims to improve computational efficiency and stability in such scenarios. While more complex task sets (e.g., simultaneous variation of multiple physical parameters) can be envisaged, their practicality warrants further discussion. For simplicity, we denote the Hamiltonian for the physical parameter $\bm{\lambda}_k$ as $\hat{H}_k \equiv \hat{H}(\bm{\lambda}_k)$ hereafter. Similarly, we simplify the variational parameters as $\theta_k \equiv \theta(\bm{\lambda}_k)$.

This curriculum learning algorithm is presented in Algorithm~\ref{alg:algorithm}. Initially, pre-training is performed on a relatively simple task, $\mathcal{T}(\bm{\lambda}_{1})$, which can be determined based on physical intuition. For instance, the computational complexity of quantum many-body systems can be determined using quantitative metrics such as the V-Score~\cite{doi:10.1126/science.adg9774}. Subsequently, efficient transfer learning $\mathcal{F}(\bm{\lambda}_{k-1}\rightarrow\bm{\lambda}_{k})$ is iteratively performed by a proper order. This iterative process of transfer learning is analogous to curriculum learning~\cite{bengio2009curriculum}, a known method for improving the stability and efficiency of learning, effects which we expect to be similarly observed in our algorithm. The algorithm is employed to exhaustively explore the task set, starting from a simple initial task and encompassing even challenging, strongly correlated regimes.

A key issue of our approach is how to determine the order of parameters $\{\bm{\lambda}_k\}_{k=1}^{K}$. In the next section, we propose a novel physical perspective to address this challenge.

\section{Perturbation Theory Perspective}
\label{5}

This section demonstrates that the above transfer learning (or curriculum learning with $K=2$) can be interpreted as perturbation theory. This interpretation reveals that transfer learning constitutes a rigorous numerical method grounded in the fundamental principles of renormalization in theoretical physics. Furthermore, we show that the energy expectation obtained via zero-shot learning for NQS is equivalent to the first-order approximation in perturbation theory. We also demonstrate that transfer learning for NQS serves as a numerical method for calculating higher-order terms in the perturbation expansion. Finally, we show that this connection to perturbation theory provides a guideline to determine the task execution order in the curriculum learning.

\subsection{Transfer Learning as Perturbation Method}
\label{5-1}

In quantum mechanics, perturbation theory provides a theoretical framework for describing how quantum states and physical quantities change in response to perturbations in the Hamiltonian (See~\ref{Appendix-QPT}). 

First, we consider the case where a perturbation of the Hamiltonian arises due to a change in the physical parameters $\bm{\nu}_k \equiv \bm{\lambda}_k - \bm{\lambda}_{k-1}$:
\begin{equation}
\hat{V}_k \equiv \hat{H}_k - \hat{H}_{k-1}.
\label{eq:perturbationH}
\end{equation}
The pre-training task is reinterpreted as the task of obtaining the NQS $|\Psi_{\theta_{k-1}}\rangle$ for the unperturbed Hamiltonian $\hat{H}_{k-1}$, where $\mathcal{T}(\bm{\lambda}_{k-1})$ is assumed to be relatively easily solvable and corresponds to the ``easy" task in transfer learning. The target task for fine-tuning can then be interpreted as the task of obtaining the NQS for the perturbed Hamiltonian $\hat{H}_k$. Generally, the perturbed Hamiltonian is more difficult to solve than the unperturbed one. This allows transfer learning to be reinterpreted as a process of ``renormalization" in the change in the wave function $|\delta\Psi_{\theta_k}\rangle$ due to the influence of the perturbation into the unperturbed state $|\Psi_{\theta_{k-1}}\rangle$.

\subsection{First Order Perturbation Energy from Zero-Shot Learning}
\label{5-2}

Here, we provide a physical interpretation of the initial value of the energy expectation during fine-tuning (which corresponds to the zero-shot prediction in the context of transfer learning) by framing the transfer learning of quantum many-body systems within the framework of perturbation theory.

First, the eigenvalue equation of Eq.~(\ref{eq:eigen}) corresponding to the pre-training task $\mathcal{T}(\bm{\lambda}_{k-1})$ can be written as:
\begin{equation}
\hat{H}_{k-1} |\Psi_{\theta_{k-1}}\rangle = E_{\theta_{k-1}} |\Psi_{\theta_{k-1}}\rangle,
\end{equation}
where $|\Psi_{\theta_{k-1}}\rangle$ and $E_{\theta_{k-1}}$ are the ground-state NQS and energy for the unperturbed Hamiltonian, respectively. In fine-tuning $\mathcal{F}(\bm{\lambda}_{k-1} \to \bm{\lambda}_k)$, the initial value of this NQS is set to $|\Psi_{\theta^{(0)}_k}\rangle \equiv |\Psi_{\theta_{k-1}}\rangle$. In this case, the energy expectation value at zero-shot prediction, $E_{\theta^{(0)}_k}$, is given by:\vspace*{-2mm}
\begin{equation}
E_{\theta^{(0)}_{k}} = E_{\theta_{k-1}} + \frac{\langle \Psi_{\theta_{k-1}} | \hat{V}_k | \Psi_{\theta_{k-1}} \rangle}{\langle \Psi_{\theta_{k-1}} | \Psi_{\theta_{k-1}} \rangle},
\label{eq:1st-perturbation}
\end{equation}
where the second term on the right-hand side corresponds to the first-order perturbation energy in perturbation theory (See~\ref{Appendix-QPT}). Thus, zero-shot prediction renormalizes the effects of first-order perturbation. In physics, it is known that for simple systems with small perturbations, the first-order perturbation provides a sufficiently accurate approximation. This corresponds to the pre-trained NQS exhibiting sufficient generalization capability for the target task. Therefore, we can introduce the accuracy of the first-order perturbation approximation $E_{\theta^{(0)}_k}$ for energy as a measure for the model generalization performance. 

Furthermore, this interpretation allows us to view the process as renormalizing the effects of higher-order (second-order and above) perturbations. In general, higher-order perturbations must be considered when dealing with complex models, but their analytical calculation is often intractable. While recent work has proposed a method for numerically calculating the effects of second-order perturbation using continuous-time quantum Monte Carlo~\cite{curry2023second}, it cannot be extended to higher-order terms. In contrast, our transfer learning approach using NQS provides a method for numerical renormalization, in principle, in all higher-order perturbation terms through fine-tuning. This interpretation also illustrates that our method clearly goes beyond the conventional scope of perturbation theory.

\subsection{Curriculum Rule}
\label{5-3}

Now, we establish a rule to determine the parameter order in the curriculum learning framework for efficient exploration of the diverse parameter space $\{\bm{\lambda}_k\}$ leveraging the aforementioned interpretation based on perturbation theory.

\begin{figure}[t]
\centering
\includegraphics[width=.5\linewidth]{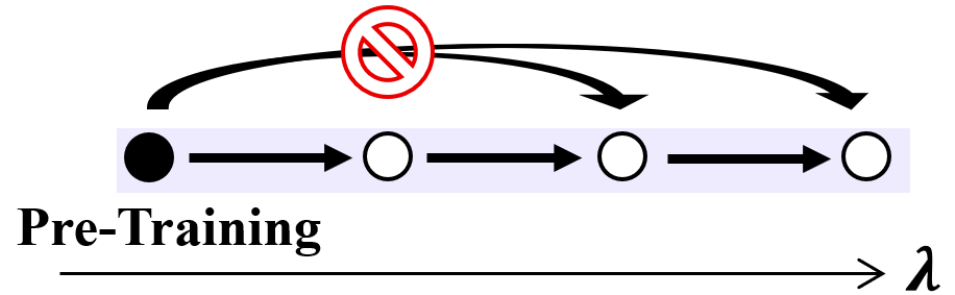}
\caption{Illustration of the parameter order in curriculum learning. Circles and the filled one represent possible targets and the pre-training task, respectively, in the parameter space of $\bm{\lambda}$. This illustrates that iteratively performing transfer learning to the nearest neighboring parameter in the space is crucial for efficient and stable curriculum learning.}
\label{fig:hubbardCL}
\end{figure}

Consider a set of parameters $\{\bm{\lambda}_k\}_{k=1}^{K}$. The curriculum order is defined as $\bm{\lambda}_1, \bm{\lambda}_2, \cdots, \bm{\lambda}_K$. The order of the parameter set can be determined based on the perturbation theory interpretation as follows. Considering $\bm{\lambda}_{k-1}$ and $\bm{\lambda}_{k}$ ($k>1$), the difference in their parameters is given by $\bm{\nu}_{k} = \bm{\lambda}_{k} - \bm{\lambda}_{k-1}$. $\bm{\nu}_k$ induces a perturbation term $\hat{V}_k$ in Eq.~(\ref{eq:perturbationH}). In perturbation theory, it is assumed that a smaller parameter change results in a smaller perturbation term $\hat{V}_k$. Therefore, $|\bm{\nu}_{k}|\ll1$ implies that fine-tuning to correct the deviation $|\delta\Psi_{k}\rangle$ becomes easier. Using this observation, the criterion to determine the order of $K-1$ parameters from the initial parameter $\bm{\lambda}_1$ is given by:
\begin{equation}
    |\bm{\lambda}_{k}-\bm{\lambda}_{k\pm1}| \le |\bm{\lambda}_{k}-\bm{\lambda}_{l}|\,\,(l\neq k,k\pm1,\,\,k>1).
    \label{eq:rule}
\end{equation}
This criterion indicates that selecting parameters such that transfer learning proceeds to the nearest neighboring parameter is optimal (cf.  Fig.~\ref{fig:hubbardCL}). Since we assume a single-axis variation of a physically meaningful parameter set as described in Section~\ref{4-2}, this criterion with the L1 norm properly determines the task order. We remark that, although Eq.~(\ref{eq:rule}) is derived under a perturbative assumption, empirical evidence shows that the same nearest-neighbor curriculum remains effective even in near‑critical regimes when the parameter increment is sufficiently small; see ~\ref{J1J2}.

Thus, our approach based on this curriculum rule can be regarded as achieving significant improvements in computational efficiency and stability by integrating the two techniques commonly employed for strongly correlated electron systems, i.e., the VMC method and the perturbation method.

\section{Pairing-Net}
\label{6}

In this section, we introduce Pairing-Net, a transferable NQS architecture capable of performing transfer and curriculum learning for strongly correlated electron systems.

\begin{figure}[t]
\centering
\includegraphics[width=.7\linewidth]{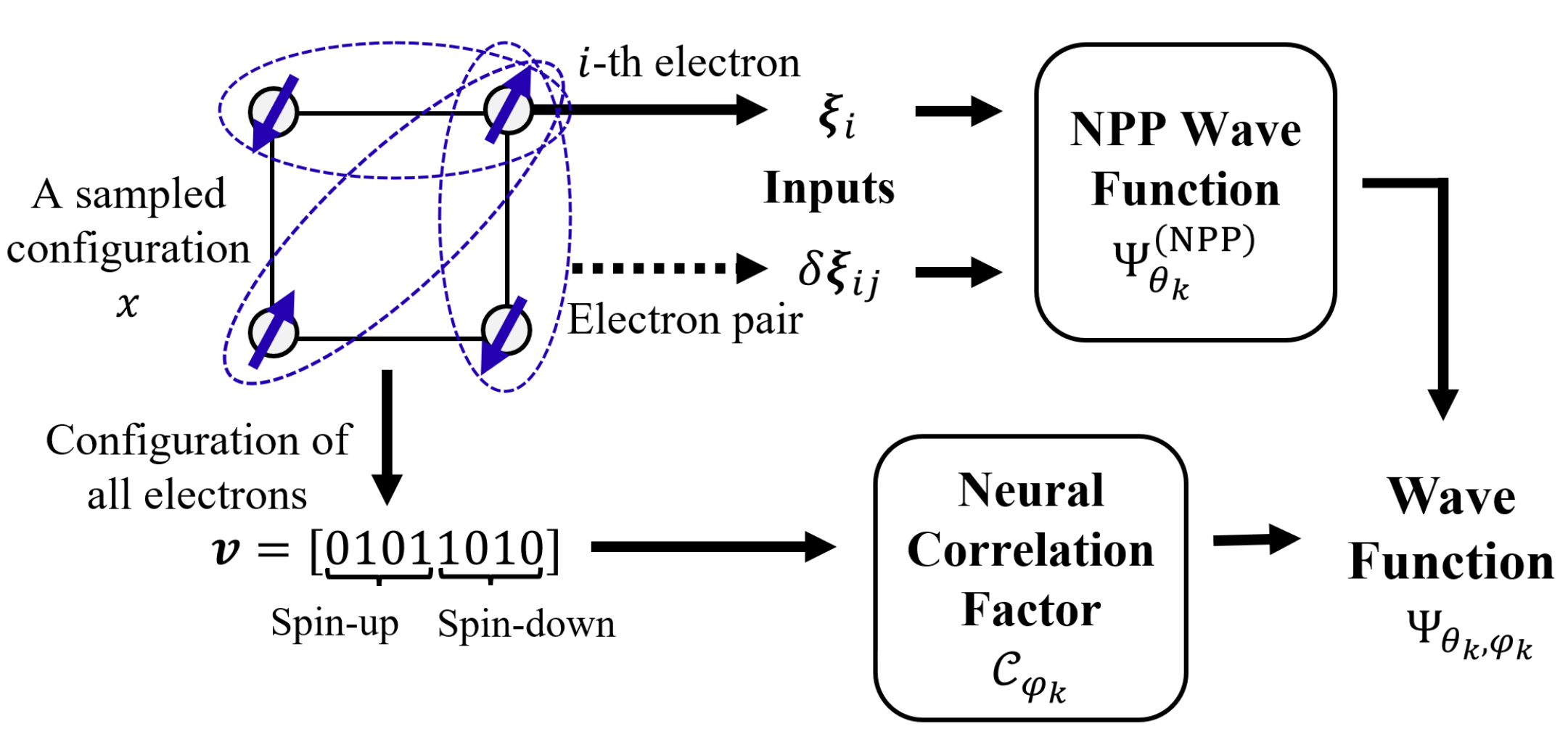}
\caption{Pairing-Net architecture. The input to the network is generated from a single sampled configuration, and the output is the wave function amplitude $\Psi_{\theta_{k},\varphi_{k}}(x) $, parameterized by neural networks $\theta_k$ and $\varphi_k$. Pairing-Net decomposes the wave function into two neural networks: A neural pair-product (NPP) wave function $\Psi^{(\mathrm{NPP})}_{\theta_{k}}(x)$ and a neural correlation factor (NCF) $C_{\varphi_{k}}(x)$. The input to the NPP wave function consists of the information $\bm{\xi}_{i}$ for each sampled configuration $x$, the information $\delta\bm{\xi}_{ij}$ for electron pairs, and the electron correlation parameters $\bm{\lambda}_{k}$ characterizing the system. The input to the NCF is a sampled configuration of all electrons $\bm{v}$.}
\label{fig:tfnqs}
\end{figure}

Existing approaches require independent training of NQS for each of the $K$ tasks in the given task set $\{\mathcal{T}(\bm{\lambda}_{k})\}_{k=1}^{K}$. For the Hubbard model, this corresponds to exploring $K$ distinct electron correlation regimes controlled by $\{\bm{\lambda}_{k}\}_{k=1}^{K}$. Independently computing these tasks for exhaustive exploration is computationally prohibitive. Pairing-Net overcomes these challenges by employing an NQS architecture designed to enable the transfer of electron correlation knowledge from one physical parameter set $\bm{\lambda}_{k-1}$ to another $\bm{\lambda}_{k}$ (Fig.~\ref{fig:tfnqs}). We assume here that the lattice size is fixed at $M_k = M$ ($k=1, \dots, K$).

First, it is empirically known that trial wave functions for strongly correlated electron systems require two key components~\cite{nomura2017restricted}: (i) antisymmetry of the wave function with respect to particle exchange of fermions, and (ii) a correction factor to accurately capture the effects of electron correlation. Therefore, Pairing-Net introduces two independent neural network architectures, a ``neural pair product (NPP) state" and a ``neural correlation factor (NCF)," to address these two requirements.

The NPP state is a neural network representation of the pair-product (PP) state~\cite{nomura2017restricted}. PP states are known to be more effective than those based on Slater determinants~\cite{pfau2020ab,hermann2020deep} for electron and spin systems (\ref{Appendix-SD} also proposes a transferable NQS using Slater determinants). Intuitively, the PP state represents a product state of electron pairs (referred to as geminals in quantum chemistry) and has the flexible expressive ability to describe, for instance, the mean-field approximation for superconductivity. The NPP state employs a neural network architecture that endows the PP state with enhanced generalization capabilities concerning the information of electron pairs. 

The input to the neural pairing function $f_{\theta_{k}}$ of the NPP state comprises the spatial coordinates and spin information $\bm{\xi}_{i} = (\bm{r}_{i}, \sigma_{i})$ for each electron at each basis, the relative spatial coordinates and spin information for electron pairs $\delta\bm{\xi}_{ij} = (\bm{r}_{i} - \bm{r}_{j}, |\bm{r}_{i} - \bm{r}_{j}|, \sigma_{i} - \sigma_{j}, |\sigma_{i} - \sigma_{j}|)$. The input dimension of $f_{\theta_{k}}$ is thus fixed, independent of the value of $\bm{\lambda}_{k}$, enabling transfer learning $\mathcal{F}(\bm{\lambda}_{k-1} \to \bm{\lambda}_{k})$. The NPP state is then expressed as follows:
\begin{equation}
    |\Psi_{\theta_{k}}^{(\mathrm{NPP})}\rangle=\left[ 
\sum_{i,j}f_{\theta_{k}}(\bm{\xi}_{i}, \delta\bm{\xi}_{ij})\hat{c}_{\bm{\xi}_{i}}^{\dagger}\hat{c}_{\bm{\xi}_{j}}^{\dagger} \right]^{N_{k}/2}|0\rangle,
\end{equation}
where $|0\rangle$ represents the vacuum state, corresponding to the absence of particles. This NPP state is capable of learning the knowledge concerning electron pairing in the system. The learned $f_{\theta_{k-1}}$ can then be transferred to different parameters $\bm{\lambda}_{k}$ through fine-tuning $\mathcal{F}(\bm{\lambda}_{k-1} \rightarrow \bm{\lambda}_{k})$. 

On the other hand, while several NQS architectures similar to the NPP state have been proposed, they are not designed for systems exhibiting complex many-body effects such as those found in the Hubbard model~\cite{gao2024neural,kim2024neural}. Therefore, to represent strongly correlated electron systems, architectural improvements are necessary to more accurately capture the effects of electron correlation.

To accurately capture electron correlation effects, we introduce a neural correlation factor (NCF). Conventionally, empirical correction factors, such as Jastrow factors~\cite{gaudin1971jastrow}, have been employed to accurately capture electron correlation effects controlled by $\bm{\lambda}_{k}$. Recently, it has been shown that representing these correlation factors with neural networks is effective even for strongly correlated electron systems~\cite{nomura2017restricted}. The NCF $C_{\varphi_k}$$:\mathbb{R}^{2M}$$\to\mathbb{R}$ takes as input the real-space electron configuration vector $\bm{v}\in\mathbb{R}^{2M}$ for each sampled configuration $x$. Since this incorporates all real-space electron configurations as input, it can capture the complex entanglement that can arise in strongly correlated electron systems. For Hubbard models with the same lattice size, the input structure remains unchanged, resulting in a transferable neural network architecture across all electron correlation regimes.

Therefore, Pairing-Net represents the trial wave function as the product of the NPP wave function $\Psi_{\theta_{k}}^{(\mathrm{NPP})}(x)\equiv\langle x|\Psi^{(\mathrm{NPP})}_{\theta_{k}}\rangle$ (For details on this calculation, see~\ref{Appendix-Pf}) and the NCF:
\begin{equation}
    \Psi_{\theta_{k},\varphi_{k}}(x) = \Psi^{(\mathrm{NPP})}_{\theta_{k}}(x)\times\exp\left[\mathcal{C}_{\varphi_{k}}(x)\right].
\end{equation}
Pairing-Net rigorously satisfies the antisymmetry requirement for fermions and constitutes NQS applicable to strongly correlated electron systems. Utilizing Pairing-Net enables efficient knowledge transfer to different systems via fine-tuning $\mathcal{F}(\bm{\lambda}_{k-1} \rightarrow \bm{\lambda}_{k})$.

\vspace*{-2.5mm}
\paragraph{Limitations:}

This study primarily focuses on benchmarking the effectiveness of curriculum learning, and thus employs a simple MLP as the neural network in Pairing-Net throughout this work (although various architectures can be employed here). From a computational accuracy perspective, using a more sophisticated neural network would be preferable. This suggests the potential for improving computational accuracy by using more advanced architectures (we experimentally confirm the effectiveness of our method even with different architectures in~\ref{Appendix-SD} and~\ref{J1J2}). For example, employing a Vision Transformer (ViT) architecture, which achieves state-of-the-art computational accuracy for frustrated spin systems~\cite{viteritti2023transformer0,rende2024simple}, for the NCF is highly promising for greater accuracy. In addition, another method that can be seamlessly integrated into this architecture is the neural backflow approach~\cite{PhysRevLett.122.226401}. In order to accurately capture the more intricate electron correlations in strongly correlated electron systems, the integration of these techniques is promising. Furthermore, modifying the architecture readily allows extension to more challenging models, such as the multiorbital Hubbard model (see~\ref{Appendix-Hubbard}), and various lattice geometries, making our method applicable to a diverse range of quantum many-body systems.

\section{Experimental Results}
\label{7}

This section presents several numerical experiments demonstrating the effectiveness of our method, including benchmark evaluations and practical applications for two-dimensional Hubbard models on a square lattice with periodic boundary conditions. First, we demonstrate that transfer learning using Pairing-Net improves computational efficiency and stability.
Next, we demonstrate that the curriculum learning facilitates the acquisition of generalization capabilities concerning electron correlation. We further demonstrate that the curriculum learning enables efficient and stable exploration of diverse electron correlation regimes.
The detailed experimental settings for this section are provided in~\ref{Appendix-EXset}.

For the $M=2 \times 2$ Hubbard model, we can use the ground-state energy obtained from exact diagonalization, $E_{\mathrm{ED}}$, as the ground truth $E_{g}$, and employ the relative error in the energy expectation value of the trained NQS, $\Delta E \equiv (E_{\theta_{k},\varphi_{k}}-E_{g})/|E_{g}|$, as a metric for convergence.
In the case of $M=4\times4$, we instead adopt the average ground-state energy, $E_{\mathrm{QMC}}$ calculated using auxiliary-field quantum Monte Carlo (AF-QMC)~\cite{qin2016benchmark}.

Furthermore, in the case of $M=2\times2$, the number of bases with $N_B=70$, all states are used directly in optimizing Eq.~(\ref{eq:variational}). Conversely, for $M=4\times4$, the number of bases at half-filling is $N_B\sim10^{14}$, clearly necessitating the use of MCMC in Eq.~(\ref{eq:mcmc}). 

\subsection{Evaluation of Transfer Learning}
\label{7-1}

\begin{table}[t]
    \centering
    \caption{Transfer learning results for $M=2\times2$ from $\tilde{U}_{1}$ to $\tilde{U}_{2}$ using Pairing-Net. The table shows the mean and standard deviation (std.) of the number of epochs to convergence, as well as the convergence count. The convergence count indicates the number of independent VMC runs (from a total of 1,000 trials) that satisfied the convergence criterion ($\Delta E < 10^{-4}$) within 500 epochs.}\vspace*{0.05in}
    \begin{tabular}{ll|rr}
         \multicolumn{1}{c}{$\tilde{U}_{1}$} & \multicolumn{1}{c}{$\tilde{U}_{2}$} & \multicolumn{1}{c}{$\mathrm{Mean}\pm\mathrm{Std.}$} & \multicolumn{1}{c}{Conv. Count} \\
        \midrule
        N/A & $2.0$ & $61 \pm 95$ & $19$ \\
        $4.0$ & $2.0$ & $2.0 \pm 0.0$ & $1000$ \\
        $6.0$ & $2.0$ & $4.0 \pm 0.0$ & $1000$ \\
        $8.0$ & $2.0$ & $5.0 \pm 0.0$ & $1000$ \\
        \midrule
        N/A & $4.0$ & $82 \pm 26$ & $70$ \\
        $2.0$ & $4.0$ & $2.0 \pm 0.0$ & $1000$ \\
        $6.0$ & $4.0$ & $2.0 \pm 0.0$ & $1000$ \\
        $8.0$ & $4.0$ & $3.0 \pm 0.0$ & $1000$ \\
    \end{tabular}
    \label{tab:tl-M4}
\end{table}

Here, we present benchmark results demonstrating that transfer learning with Pairing-Net significantly improves both computational efficiency and stability. Here, we present benchmark results demonstrating that transfer learning with Pairing-Net significantly improves both computational efficiency and stability compared to the conventional VMC approach that does not employ transfer learning.

Table~\ref{tab:tl-M4} presents benchmark results for $M=2\times2$ demonstrating transfer learning with Pairing-Net. Results where pre-training is listed as N/A correspond to the conventional approach using random initialization. 
For each transfer learning instance, we performed 1000 independent VMC calculations, and the table shows the mean and standard deviation (std.) of the number of epochs required for convergence.
Here we define convergence as $\Delta E<1.0\times10^{-4}$ (see~\ref{Appendix-pretrained} for the impact of pre-trained NQS accuracy on transfer learning).
The convergence count, defined as the number of independent VMC calculations (out of 1,000 trials) that achieved $\Delta E < 10^{-4}$ within a maximum of 500 epochs, is also shown.
The mean number of epochs to convergence serves as a metric for efficiency, while the std. of the number of epochs to convergence and the convergence count are used as metrics for stability.
The results demonstrate that transfer learning achieves significant computational speedups compared to the conventional approach.
For instance, transfer learning from $\tilde{U}_{1} = 2.0$ to $\tilde{U}_{2} = 4.0$ shows approximately a $40$-fold speedup compared to the conventional approach.
Furthermore, the results for the std. in the number of epochs to convergence and the convergence rate demonstrate that transfer learning significantly improves the stability of the computation across all regimes. While the conventional approach exhibits large std. in the number of epochs to convergence, all instances using pre-trained NQS show zero std. Moreover, the conventional approach with $\tilde{U}_2 = 2.0$ exhibits a very low convergence rate of $1.9\%$, indicating significant instability, whereas transfer learning achieves $100\%$ convergence.

\begin{figure}[t]
\centering
\includegraphics[width=.7\linewidth]{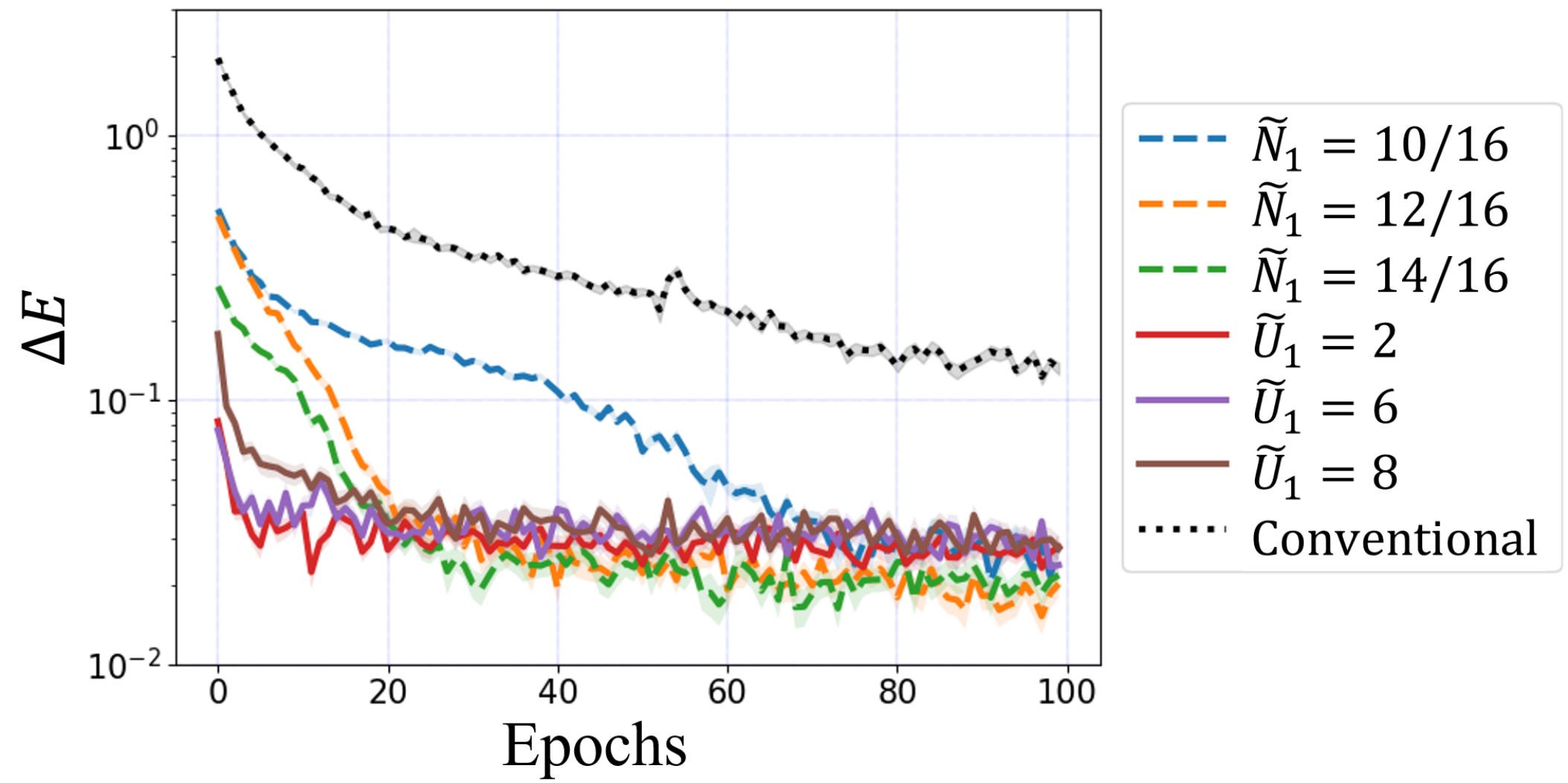}
\caption{The loss curves for $M=4\times4$ over 100 epochs applied to $(\tilde{N}_{2},\tilde{U}_{2})=(16/16,4)$ using the various pre-trained NQS.}
\label{fig:M16-NUloss}
\end{figure}

Next, we present benchmark results for the $M=4\times4$ square-lattice Hubbard model. While existing multi-task learning methods~\cite{zhu2023hubbardnet} that do not rely on MCMC are impractical for this scale of the Hubbard model due to computational cost and instability issues, we demonstrate the scalability and effectiveness of our approach below.

Fig.~\ref{fig:M16-NUloss} presents the results of transfer learning for $M=4 \times 4$ with $\lambda_{k}=(\tilde{N}_{k},\tilde{U}_{k})$. First, considering the results for $\tilde{U}_{k}$, significant efficiency improvements are observed compared to the conventional approach. For instance, when using a pre-trained NQS with $\tilde{U}_{1}=2.0$, it requires 301 epochs to reach an initial value of $\Delta E^{(0)}=8.3\times10^{-2}$ (see Table~\ref{tab:LSM16} in~\ref{Appendix-LS}) with the conventional approach (see~\ref{Appendix-M16-stab} for details on stability with MCMC).
Next, examining the results for the electron density $\tilde{N}_{k}$, although the improvements are less pronounced than those for $\tilde{U}_{k}$, substantial acceleration is still achieved. Even more surprisingly, we observe a tendency for higher accuracy when transfer learning is applied to the electron density (the trend was also confirmed for the larger $M = 6 \times 6$; see~\ref{Appendix-MoreLS}).

Although it is non-trivial whether the perturbative intuition-based rule in Eq.~(\ref{eq:rule}) holds in typical non-perturbative regions, including strong coupling regimes ($\tilde{U}\gtrsim6$), remarkably, these results strongly suggest that the rule remains valid in such scenarios. This robustness can be attributed to the non-perturbative, VMC-based fine-tuning approach; notably, we show that the computational efficiency is maintained even when crossing a phase boundary in~\ref{J1J2}.
Furthermore, it is important to note that our VMC-based method is not directly hampered by the fermionic sign problem, making it applicable to a broader range of electron densities compared to the AF-QMC method.
These results demonstrate that even for large-scale systems requiring MCMC, our method remains effective with high stability.

\begin{figure}[t]
\centering
\includegraphics[width=.6\linewidth]{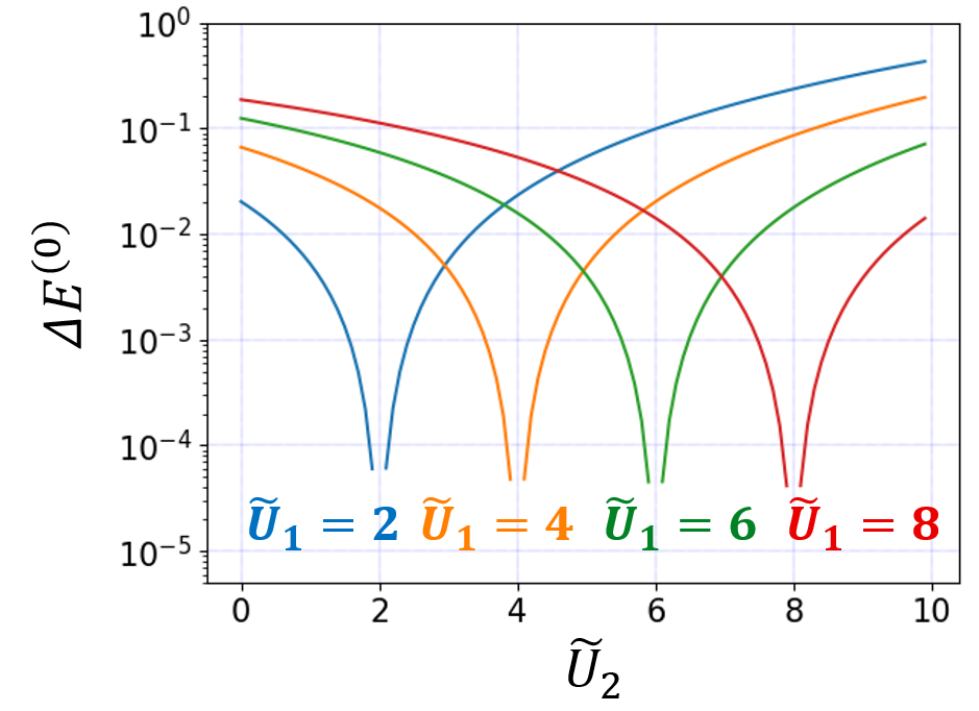}
\caption{The dependence of $\Delta E^{(0)}$, acquired through pre-training, on changes in the target parameter $\tilde{U}_2$. Pre-trained NQS with $\tilde{U}_1 = 2, 4, 6, 8$ were used.}
\label{fig:generalization}
\end{figure}

\subsection{Generalization of Electron Correlations}
\label{7-2}

Here, we present benchmark results on the generalization performance of electron correlation effects acquired through pre-training with Pairing-Net. Based on the theoretical analysis in Section~\ref{5-2}, we utilize the accuracy of the first-order perturbed energy, $E_{\theta^{(0)}_2, \varphi^{(0)}_2}$, as a metric for generalization performance. In our experiments, the generalization performance of the pre-trained NQS is evaluated using the relative error $\Delta E^{(0)} \equiv (E_{\theta^{(0)}_2, \varphi^{(0)}_2} - E_{g})/|E_{g}|$. Here, $\Delta E^{(0)}$ directly represents the accuracy of the first-order perturbation energy approximation; thus, lower values of $\Delta E^{(0)}$ indicate higher generalization performance.

Fig.~\ref{fig:generalization} shows how the generalization performance acquired during pre-training behaves as a function of the parameter change, $\nu_{2}=\tilde{U}_{2} - \tilde{U}_{1}$. Across all pre-trained NQS, higher generalization performance ($\Delta E^{(0)}$) is consistently observed. This trend, specifically the observed behavior in fine-tuning with small parameter changes, has also been empirically observed in the $J_1$-$J_2$ Heisenberg model, a prototypical frustrated spin system~\cite{rende2024fine}. 
The results support the theoretical analysis in Section~\ref{5-3}, indicating that curriculum learning with incrementally increasing, small $\nu_{2}$ values facilitates learning that leverages both high generalization performance and optimization stability.

\subsection{Explorative Curriculum Learning}
\label{7-3}

Here we examine the total number of epochs required for transfer learning during curriculum learning. In this experiment, we consider a challenging task set $\tilde{U}_{k} = 2.0, 4.0, 6.0, 8.0$, frequently adopted to investigate intriguing behaviors of the Hubbard model~\cite{qin2016benchmark,nomura2017restricted,PhysRevB.99.195141,ijms24021509}. Such task sets correspond, for example, to analyzing various crystal structure deformation patterns in real materials~\cite{PhysRevB.99.195141,ijms24021509}. Table~\ref{tab:cl-m16} presents the total number of epochs excluding pre-training for the conventional approach (no curriculum learning, indicated as N/A) and for several curricula; see~\ref{Appendix-DRCL} for 
the details of the results.
The efficiency improvement is greater for the larger $M=4 \times 4$ system compared to $M=2 \times 2$, suggesting that the curriculum following our proposed rule becomes increasingly advantageous for larger system sizes.
For the case of $M=4\times4$, our curriculum achieves up to a 15-fold speedup compared to a random curriculum and an over 200-fold speedup compared to the conventional approach.
While this analysis is limited to only four tasks, we anticipate even greater efficiency gains with our method as the search space expands.

\begin{table}[t]
    \centering
    \caption{Results for several curricula applied to the task set with $\tilde{U}_{k} = 2.0, 4.0, 6.0, 8.0$, starting from $\tilde{U}_1 = 2.0$. The total number of epochs for $M=2\times2$ and $4\times4$ to convergence is shown.}
    \vspace*{0.1in}
    \begin{tabular}{l|rr}
        \multicolumn{1}{c}{Curriculum} & \multicolumn{1}{c}{$2\times2$} & \multicolumn{1}{c}{$4\times4$} \\
        \midrule
        N/A & $246$ & $792$  \\
        Random & $11$ & $45$ \\
        Ours & $7$ & $3$ 
    \end{tabular}
    \label{tab:cl-m16}
\end{table}

\section{Conclusions}
\label{8}

In this study, we propose a novel curriculum learning framework for the efficient and stable exploration of a broad parameter space in quantum many-body systems. The framework introduces transfer learning for NQS and subsequently proposes a novel curriculum learning algorithm based on iterative transfer learning. By recognizing the interpretability of this transfer learning process through the lens of perturbation theory, we provide guidelines for curriculum design in quantum many-body systems. Furthermore, we introduce Pairing-Net, a transferable NQS architecture designed for the practical application of this curriculum learning approach to strongly correlated electron systems, and demonstrate its effectiveness through numerical experiments. These experiments show a speedup of approximately $200$-fold or more compared to existing approaches, along with a significant improvement in computational stability. These findings suggest that our proposed method has the potential to significantly mitigate the computational cost and instability issues inherent in existing methods.

\section*{Reproducibility}

The implementation code used to generate all results in this paper is available from the authors upon reasonable request.

\ack

This study was supported by JST CREST Grant Number JPMJCR1913 and JSPS KAKENHI Grant Number JP22H00516, JP22H05106 (Y.K.), and JP25K15230 (T.K.).

\section*{Data availability statement}

No new datasets were generated or analyzed in the course of this study.

\newpage
\appendix

\section{Hubbard Model}
\label{Appendix-Hubbard}

In this section, we introduce the intuitive physical significance of the Hubbard model. Furthermore, we explain the multiorbital Hubbard model, which becomes essential when quantitatively considering real materials, and demonstrate that its parameter space is significantly larger compared to that of the single-orbital model in Eq.~(\ref{eq:hubbard}).

\begin{figure}[t]
\centering
\includegraphics[width=.5\linewidth]{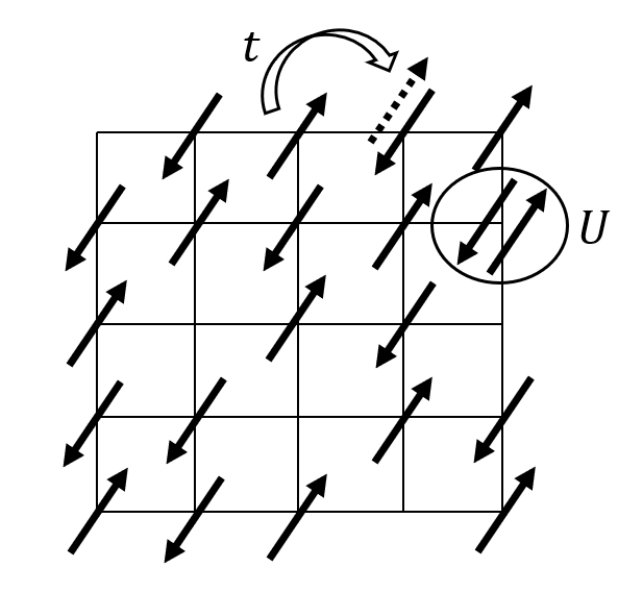}
\caption{Schematic representation of the two-dimensional square lattice Hubbard model. The arrows indicate electrons on the lattice with spin-up or spin-down. $t$ denotes the electron hopping between nearest-neighbor lattice sites, and $U$ represents the on-site Coulomb repulsion.}
\label{fig:HubbardModel}
\end{figure}

The Hubbard model has been extensively studied as a theoretical framework for understanding magnetic and superconducting materials. In this model, the electronic states in a crystalline solid are described by considering only two main effects: The kinetic term denoted as $\hat{H}_{\mathrm{Kin}} \equiv -t \sum_{\langle i,j \rangle} \sum_{\sigma} \left( \hat{c}_{i\sigma}^{\dagger} \hat{c}_{j\sigma} + \text{H.c.} \right)$, and the interaction term, representing the repulsive Coulomb interaction between electrons denoted as $\hat{H}_{\mathrm{int}} \equiv U \sum_{i=1}^N \hat{n}_{i\uparrow} \hat{n}_{i\downarrow}$. These effects are illustrated in Fig.~\ref{fig:HubbardModel}.

Although the Hubbard model is a highly simplified model, it has been shown to exhibit a wide range of physical phenomena, such as magnetism and superconductivity, making it an intriguing and widely studied model~\cite{arovas2022hubbard}. These phenomena are believed to arise from electron correlations, which also present significant challenges in understanding the model.
The balance between kinetic and interaction terms governs electron behavior: the kinetic term enhances itinerancy, while the interaction term induces localization.
The degree of electron correlation is determined by parameters such as the electron density $\tilde{N}$ and the ratio $\tilde{U} = U/t$. Depending on these values, the system can exhibit metallic, insulating, or even superconducting behavior, as suggested by numerous studies~\cite{hirsch1985two,anderson1997theory,imada1998rev}. Thus, exhaustively exploring quantum states across various parameter sets $\bm{\lambda} = (\tilde{N}, \tilde{U})$ is crucial for analyzing these phenomena. 

Moreover, accurately modeling realistic materials requires extending the Hubbard model to incorporate multiorbital effects to capture the essential physics, e.g., iron-based and unconventional cuprate superconductors~\cite{PhysRevLett.101.087004,PhysRevB.99.224515}. The effective Hamiltonian of the multiorbital Hubbard model can also be divided into the kinetic term, $\hat{H}_{\mathrm{Kin}}^{(\mathrm{MO})}$, describing electron hopping, and the interaction term, $\hat{H}_{\mathrm{int}}^{(\mathrm{MO})}$, describing electron-electron interactions. These terms can be expressed as follows:
\begin{align}
    \hat{H}_{\mathrm{Kin}}^{(\mathrm{MO})}&=\sum_{\langle i,j \rangle}\sum_{\zeta,\eta}\sum_{\sigma} (t_{ij}^{\zeta\eta}\hat{c}_{i\zeta\sigma}^{\dagger}\hat{c}_{j\eta\sigma}+\mathrm{H.c.})+\sum_{i,\zeta,\sigma}\epsilon_{\zeta}\hat{n}_{i\zeta\sigma},\\
    \hat{H}_{\mathrm{int}}^{(\mathrm{MO})}&=U\sum_{i,\zeta}\hat{n}_{i\zeta\uparrow}\hat{n}_{i\zeta\downarrow}+U'\sum_{i,\zeta,\eta}\sum_{\sigma.\tau}\hat{n}_{i\zeta\sigma}\hat{n}_{i\eta\tau} \notag\\ 
    &\quad-J\sum_{i,\zeta,\eta}\sum_{\sigma,\tau}\hat{c}_{i\zeta\sigma}^{\dagger}\hat{c}_{i\zeta\tau}\hat{c}_{i\eta\tau}^{\dagger}\hat{c}_{i\eta\sigma} \notag\\
    &\quad+J'\sum_{i,\zeta\neq\eta}\hat{c}_{i\zeta\uparrow}^{\dagger}\hat{c}_{i\zeta\downarrow}^{\dagger}\hat{c}_{i\eta\downarrow}\hat{c}_{i\eta\uparrow},
\end{align}
where $\hat{c}^{\dagger}_{i\zeta\sigma}$ ($\hat{c}_{i\zeta\sigma}$) is the creation (annihilation) operator for an electron with spin $\sigma$ in the $\zeta$ orbital at site $i$. Additionally, $t_{ij}^{\zeta\eta}\in\mathbb{R}$ represents the electron hopping from the $\eta$ orbital at site $j$ to the $\zeta$ orbital at site $i$, $\epsilon_{\zeta}\in\mathbb{R}$ denotes the energy level of the $\zeta$ orbital, and $\hat{n}_{i\zeta\sigma} \equiv \hat{c}_{i\zeta\sigma}^{\dagger}\hat{c}_{i\zeta\sigma}$.
Here, H.c. denotes the Hermitian conjugate of the preceding term, $t_{ji}^{\eta\zeta}\hat{c}_{j\eta\sigma}^{\dagger}\hat{c}_{i\zeta\sigma}$.
In the case of a multiorbital system, $U\in\mathbb{R}$ represents the on-site Coulomb interaction within the same orbital, $U'\in\mathbb{R}$ represents the inter-orbital on-site Coulomb interaction, $J\in\mathbb{R}$ represents the Hund's coupling, and $J'\in\mathbb{R}$ represents the pair hopping. As such, multiorbital systems possess a richer set of parameters, $\bm{\lambda}^{(\mathrm{MO})}=(\tilde{N},t_{ij}^{\zeta\eta},\epsilon_{\zeta},U,U',J,J')$, compared to the single-orbital case, $\bm{\lambda}=(\tilde{N},\tilde{U})$. Consequently, the search space is significantly larger than in the single-orbital case, making efficient exploration crucial. Our proposed method can be readily extended to multiorbital systems, suggesting its capability to efficiently explore the aforementioned vast parameter space.

\section{Perturbation Theory}
\label{Appendix-QPT}

We introduce perturbation theory, a frequently used approximation method in quantum mechanics. Perturbation theory solves for physical quantities and quantum states by expressing them as a series expansion in terms of the solution of a solvable system (unperturbed system), when the system of interest is composed of this unperturbed system and a small contribution from a perturbation. Perturbation theory is also employed in methods to improve the accuracy of the solution by refining the series expansion, and in approximation techniques that extract and sum the dominant contributions (often calculated using Feynman diagrams, hence referred to as diagrammatic methods, e.g., random phase approximation~\cite{PhysRev.82.625} and fluctuation-exchange approximation~\cite{PhysRevLett.62.961}). In what follows, we introduce the Rayleigh-Schrödinger perturbation expansion (time-independent), the most fundamental type of perturbation theory.

We consider a Hamiltonian $\hat{H}_{k}$ given by:
\begin{equation}
\hat{H}_{k} = \hat{H}_{k-1} + \hat{V}_{k},
\label{eq:perturab}
\end{equation}
where $\hat{H}_{k-1}$ is the unperturbed Hamiltonian and $\hat{V}_{k}$ is the perturbation term, both assumed to be time-independent. Let us assume that the normalized complete set of eigenstates $|\Psi_{k-1,n}\rangle$ of the unperturbed part $\hat{H}_{k-1}$ is obtained as follows:
\begin{align}
&\hat{H}_{k-1} |\Psi_{k-1,n}\rangle = E_{k-1,n} |\Psi_{k-1,n}\rangle,\\
\
&\langle \Psi_{k-1,n} | \Psi_{k-1,m} \rangle = \delta_{nm},
\end{align}
where $|\Psi_{k-1,n}\rangle$ denotes the $n$-th eigenstate of $\hat{H}_{k-1}$, corresponding to the energy eigenvalue $E_{k-1,n}$.
We consider the eigenstate $|\Psi_{k-1,n}\rangle$ without energy degeneracy among these states. 

Our goal is to solve the following eigenvalue equation:
\begin{equation}
    \hat{H}_{k}|\Psi_{k,n}\rangle=E_{k,n}|\Psi_{k,n}\rangle.
    \label{eq:perturbeigen}
\end{equation}
In the following, we derive the first-order correction term for the energy eigenvalues. 

The first-order analysis of the eigenvalues relies on the zeroth-order approximation of the corresponding eigenstates,
\begin{equation}
    |\Psi_{k,n}\rangle\approx|\Psi_{k-1,n}\rangle.
\end{equation}
Using this approximation, the energy eigenvalue $E_{k,n} = \langle \Psi_{k,n} | \hat{H}_k | \Psi_{k,n} \rangle$ can be approximated as follows:
\begin{align}
    E_{k,n}&\approx E_{k-1,n}+E_{k,n}^{(1)},\\
    E_{k,n}^{(1)}&=\langle\Psi_{k-1,n}|\hat{V_{k}}|\Psi_{k-1,n}\rangle.
\end{align}
This represents the first-order result of perturbation theory for the energy levels.
Considering the ground-state energy for $n=0$, this expression is found to be identical in form to the energy correction term for zero-shot prediction in Eq.~(\ref{eq:1st-perturbation}). While the first-order perturbation energy provides a sufficient approximation in simple cases, such as applying an external electric field to a hydrogen atom, calculating higher-order perturbation terms (second-order and above) often becomes necessary for complex systems. However, obtaining analytical solutions for these higher-order terms is challenging, necessitating numerical computation. Our fine-tuning method can be viewed as a technique for obtaining these higher-order terms.

\section{The NPP wave function with Pfaffian}
\label{Appendix-Pf}

The NPP wave function for a sampled real space configuration $x$ is expressed using the neural pairing function $f_{\theta_{k}}(\bm{\xi}_{i}, \delta\bm{\xi}_{ij})$ that incorporates information about electron pairs , as follows:
\begin{equation}
    \Psi^{(\mathrm{NPP})}_{\theta_{k}}(x) = \mathrm{Pf}[X_{\theta_{k}}].
    \label{eq:NPP}
\end{equation}
Here, a skew-symmetric matrix $X_{\theta_{k}}\in\mathbb{C}^{N_{k}\times N_{k}}$ is defined as $X_{\theta_{k}} = F_{\theta_{k}} - F_{\theta_{k}}^{\mathrm{T}}$, where $F_{\theta_{k}}\in\mathbb{C}^{N_{k}\times N_{k}}$ is the matrix representation of the neural pairing function with $(F_{\theta_k})_{ij} = f_{\theta_k}(\bm{\xi}_{i}, \delta\bm{\xi}_{ij})$. $\mathrm{Pf}$ denotes the Pfaffian, which is defined for a $2L \times 2L$ skew-symmetric matrix $A$ as follows:
\begin{equation}
\mathrm{Pf}[A] = \frac{1}{2^L L!} \sum_{\sigma \in S_{2L}} \mathrm{sgn}(\sigma) \prod_{i=1}^L A_{\sigma(2i-1), \sigma(2i)},
\label{eq:pfaffian}
\end{equation}
where $S_{2L}$ denotes the symmetric group of order $2L$.  In the case where the number of electrons is odd, the above formulation can be applied by considering the unpaired single-particle orbital~\cite{kim2024neural,gao2024neural}.

\begin{table*}[t]
    \centering\small
    \caption{Hyperparameters in our experiments for the Hubbard models.}
    \label{tab:conditions}
    \vspace*{2mm}\small
    \begin{tabular}{l|rrrr}
    \multicolumn{1}{c}{\makecell{System size \\ NQS architecture}}
    & \multicolumn{1}{c}{\makecell{$M = 2 \times 2$ \\ Pairing-Net}}
    & \multicolumn{1}{c}{\makecell{$M = 2 \times 2$ \\ SD-Net}}
    & \multicolumn{1}{c}{\makecell{$M = 4 \times 4$ \\ Pairing-Net}}
    & \multicolumn{1}{c}{\makecell{$M = 6 \times 6$ \\ Pairing-Net}}
    \\ \midrule
    Number of layers (NPP)           &    1 &    1 &    1 &     1 \\
    Hidden size (NPP)                &   50 &   50 &   50 &   100 \\
    Number of layers (NCF)           &    1 &    1 &    1 &     2 \\
    Hidden size (NCF)                &  100 &  100 &  100 &   400 \\
    Number of variational parameters & 1452 & 1452 & 3852 & 59502 \\
    Number of epochs in pre-training & 1000 & 1000 & 1000 &  1000 \\
    Number of MC samples             &    - &    - & 1024 &  1024 \\ 
    Number of independent MCs        &    - &    - &    8 &     8 \\
    Length of Markov chain           &    - &    - & 4736 &  4736
    \end{tabular}
\end{table*}

\section{Experimental Setup}
\label{Appendix-EXset}

Here, we provided the detailed settings for each experiment conducted in Section~\ref{7} and subsequent Appendices. First, the implementation of Pairing-Net, SD-Net in~\ref{Appendix-SD}, and ViT wave function in~\ref{J1J2} utilized NetKet~\cite{netket3:2022}, a powerful open-source library for quantum many-body problems. This study utilized a single NVIDIA H100 PCIe GPU with $81,559$ MiB of total memory. 

The architectures of the neural networks for the NPP wave function, NSD wave function, and NCF all employed multilayer perceptrons (MLPs), with the hyperbolic tangent function used as the activation function in all cases. All networks in Pairing-Net and SD-Net were implemented with purely real-valued parameters; in our code, the NPP/NSD wave-function modules returned a two-component output— the real and the imaginary part of the wave function. Optimization was performed using the stochastic reconfiguration (SR) method and a Metropolis-Hastings sampler implemented in NetKet, and random initialization was employed for cases without fine-tuning. The learning rate was scheduled to decrease from $0.05$ to $0.01$. The hyperparameters and computation times for the experiments are summarized in Table~\ref{tab:conditions} and~\ref{tab:time}, respectively.

Furthermore, the ViT wave function implementation used in~\ref{J1J2} was adopted from the repository at \url{https://zenodo.org/records/14060431}~\cite{Rende_2025} and the ViT wave function section of the NetKet tutorial documentation (\url{https://netket.readthedocs.io/en/latest/index.html}). The ViT wave function employed Factored Multi-Head Attention~\cite{Rende_2025} with 10 attention heads, 4 encoder layers, an embedding dimension of 60, and a patch size of $2\times2$, resulting in a total of 154,980 parameters.

\begin{table}[h!]
    \centering\small
    \caption{Computation times for 1000 epochs.}
    \vspace*{2mm}\small
    \label{tab:time}
    \begin{tabular}{ccc|r}
    Model & $M$ & NQS & time (GPU hours) \\ \hline
    Hubbard & $2 \times 2$ & Pairing-Net & $2.7\times10^{-3}$ \\ 
    Hubbard & $2 \times 2$ & SD-Net & $2.6\times10^{-3}$ \\ 
    Hubbard & $4 \times 4$ & Pairing-Net & $9.4$ \\ 
    Hubbard & $6 \times 6$ & Pairing-Net & $46$ \\ 
    $J_1$-$J_2$ Heisenberg & $6\times6$ & ViT & $4.3\times10^{-1}$
    \end{tabular}
\end{table}

\section{Detailed Results for Transfer Learning}
\label{Appendix-DRTL}

Here, we summarize various numerical experiment results for the $M=2\times2$ and $4\times4$ Hubbard models in Section~\ref{7-1}. The computational conditions are the same as those employed in Section~\ref{7-1}. Furthermore, as described in Section 7, the threshold used to define convergence is based on the order of computational accuracy achievable after 100 epochs with the conventional approach; specifically, $10^{-4}$ for the 2$\times$2 case and $10^{-1}$ for the $4\times4$ case.
The results obtained here were utilized in the curriculum learning analysis presented in Section~\ref{7-3}.

\subsection{Results of $M=2\times2$}
\label{Appendix-DRTL-Small}

Table~\ref{tab:tl-M4-Appendix} details the transfer learning results for the $M=2\times2$ case. Results where pre-training is listed as N/A represent the conventional approach. Convergence is defined as $\Delta E < 1.0 \times 10^{-4}$. The table shows the mean and standard deviation (std.) of the number of epochs to convergence, and the convergence rate, based on $1000$ independent trials. A hyphen (-) indicates failure to converge.

First, the upper table in Table~\ref{tab:tl-M4-Appendix} presents the results for $\tilde{U}_{2} = 6.0$ and $8.0$, which were not covered in Section~\ref{7-1}. These results are consistent with the theoretical analysis in Section~\ref{5-3}. As in the results of Section~\ref{7-1}, they demonstrate the effectiveness of our method for the $M=2\times2$ Hubbard model in terms of both computational efficiency and stability.

\begin{table}[t]
    \centering\small
    \caption{Transfer learning results for $M=2\times2$ under various conditions using Pairing-Net. The upper panel shows the results of transfer learning with fixed $\tilde{N}$ and varying $\tilde{U}_{k}$, while the lower panel shows the results with fixed $\tilde{U}$ and varying $\tilde{N}_{k}$.}
    \vspace*{2mm}\small
    \begin{tabular}{lll|rr}
        \multicolumn{1}{c}{$\tilde{N}$} & \multicolumn{1}{c}{$\tilde{U}_{1}$} & \multicolumn{1}{c}{$\tilde{U}_{2}$} & \multicolumn{1}{c}{$\mathrm{Mean}\pm\mathrm{Std.}$} & \multicolumn{1}{c}{Conv. Count} \\
        \midrule
        $1.0$ & N/A & $6.0$ & $83\pm15$ & $293$ \\
        $1.0$ & $2.0$ & $6.0$ & $3.0\pm0.0$ & $1000$ \\
        $1.0$ & $4.0$ & $6.0$ & $2.0\pm0.0$ & $1000$ \\
        $1.0$ & $8.0$ & $6.0$ & $2.0\pm0.0$ & $1000$ \\
        \midrule
        $1.0$ & N/A & $8.0$ & $81\pm13$ & $415$ \\
        $1.0$ & $2.0$ & $8.0$ & $5.0\pm0.0$ & $1000$ \\
        $1.0$ & $4.0$ & $8.0$ & $4.0\pm0.0$ & $1000$ \\
        $1.0$ & $6.0$ & $8.0$ & $3.0\pm0.0$ & $1000$ \\
        \multicolumn{1}{c}{}\\
        \multicolumn{1}{c}{$\tilde{U}$} & \multicolumn{1}{c}{$\tilde{N}_{1}$} & \multicolumn{1}{c}{$\tilde{N}_{2}$} & \multicolumn{1}{c}{$\mathrm{Mean}\pm\mathrm{Std.}$} & \multicolumn{1}{c}{Conv. Count} \\
        \midrule
        $2.0$ & N/A & $0.5$ & $48 \pm 22$ & $365$ \\
        $2.0$ & $1.0$ & $0.5$ & - & $0$ \\
        $2.0$ & N/A & $1.0$ & $61 \pm 95$ & $19$ \\
        $2.0$ & $0.5$ & $1.0$ & $(1.5 \pm0.0)\times 10^2 $ & $1000$ \\
        \midrule
        $4.0$ & N/A & $0.5$ & $49 \pm 22$ & $333$ \\
        $4.0$ & $1.0$ & $0.5$ & - & $0$ \\
        $4.0$ & N/A & $1.0$ & $82 \pm 26$ & $70$ \\
        $4.0$ & $0.5$ & $1.0$ & $(1.6 \pm0.0)\times 10^2$ & $1000$ \\
        \midrule
        $6.0$ & N/A & $0.5$ & $50 \pm 21$ & $317$ \\
        $6.0$ & $1.0$ & $0.5$ & $10 \pm 0.0$ & $1000$ \\
        $6.0$ & N/A & $1.0$ & $83 \pm 15$ & $293$ \\
        $6.0$ & $0.5$ & $1.0$ & $53 \pm 0.0$ & $1000$ \\
        \midrule
        $8.0$ & N/A & $0.5$ & $52 \pm 21$ & $304$ \\
        $8.0$ & $1.0$ & $0.5$ & $43 \pm 0.0$ & $1000$ \\
        $8.0$ & N/A & $1.0$ & $81 \pm 13$ & $415$ \\
        $8.0$ & $0.5$ & $1.0$ & $60 \pm 0.0$ & $1000$ \\
    \end{tabular}
    \label{tab:tl-M4-Appendix}
\end{table}

Next, the lower table in Table~\ref{tab:tl-M4-Appendix} presents the results of transfer learning for $M=2\times2$ focusing on changes in $\tilde{N}_{k}$. As shown, the convergence speed is slower compared to the results for $\tilde{U}_{k}$. This can be attributed to the significant finite-size effects in small-scale systems such as $M=2\times2$, which limit the ability to make small changes in electron density. Here, the finite-size effect refers to the fact that the relative change in electron number, $|N_{2}-N_{1}|$, becomes significantly larger in small-scale systems compared to large-scale systems with respect to the lattice size $M$. These results are consistent with the theoretical analysis in Section~\ref{5-3}. Indeed, as observed in the results for the $M=4\times4$ and $M=6\times6$ cases (Section~\ref{7-1} and~\ref{Appendix-MoreLS}, respectively), for a given change in the number of electrons, the convergence speed tends to increase with system size compared to the results presented in this section.

\begin{table}[t]
    \centering\small
    \caption{Transfer learning results for $M=4\times4$. The first section shows results for transfer learning from $\tilde{U}_{1}$ to $\tilde{U}_{2}$ with a fixed $\tilde{N}=1.0$ using Pairing-Net, while the second section presents results under a fixed $\tilde{U}=4.0$ in electron density. The table shows the number of epochs to convergence, and $\Delta E^{(0)}$. Additionally, the rightmost column $N_{E}^{(0)}$ indicates the number of epochs required for convergence to the initial value $\Delta E^{(0)}$ using the conventional approach.}
    \vspace*{2mm}\small
    \begin{tabular}{ll|rrr}
        \multicolumn{1}{c}{$\tilde{U}_{1}$} & \multicolumn{1}{c}{$\tilde{U}_2$}  & \multicolumn{1}{c}{$\Delta E^{(0)}$} & \multicolumn{1}{c}{Conv. Epochs} & $N_{E}^{(0)}$ \\
        \midrule
        N/A & $2.0$ & $1.1$ & $50$ & N/A \\
        $4.0$ & $2.0$ & $5.0\times10^{-2}$ & $0$ & 143\\
        $6.0$ & $2.0$ & $1.5\times10^{-1}$ & $1$ & 37\\
        $8.0$ & $2.0$ & $2.7\times10^{-1}$ & $2$ & 19\\
        \midrule
        $2.0$ & $4.0$ & $8.3\times10^{-2}$ & $0$ & 301\\
        N/A & $4.0$ & $2.0$ & $188$ & N/A \\
        $6.0$ & $4.0$ & $7.7\times10^{-2}$ & $0$ & 337\\
        $8.0$ & $4.0$ & $1.8\times10^{-1}$ & $1$ & 68\\
        \midrule
        $2.0$ & $6.0$ & $3.2\times10^{-1}$ & $4$ & 49\\
        $4.0$ & $6.0$ & $1.1\times10^{-1}$ & $1$ & 244\\
        N/A & $6.0$ & $2.8$ & $258$ & N/A \\
        $8.0$ & $6.0$ & $8.8\times10^{-2}$ & $0$ & 261\\
        \midrule
        $2.0$ & $8.0$ & $7.8\times10^{-1}$ & $41$ & 19\\
        $4.0$ & $8.0$ & $3.4\times10^{-1}$ & $6$ & 189\\
        $6.0$ & $8.0$ & $1.2\times10^{-1}$ & $2$ & 309\\
        N/A & $8.0$ & $5.1$ & $346$ & N/A \\
        \multicolumn{2}{c}{}\\
        \multicolumn{1}{c}{$\tilde{N}_{1}$} & \multicolumn{1}{c}{$\tilde{N}_{2}$} & \multicolumn{1}{c}{$\Delta E^{(0)}$} & \multicolumn{1}{c}{Conv. Epochs} \\
        \midrule
        $\text{N/A}$ & $1.00$ & 2.0 & $188$ & N/A \\
        $0.875$ & $1.00$ & $2.7\times10^{-1}$ & $10$ & 44\\
        $0.750$ & $1.00$ & $5.0\times10^{-1}$ & $14$ & 17\\
        $0.625$ & $1.00$ & $5.3\times10^{-1}$ & $27$ & 16\\
    \end{tabular}
    \label{tab:LSM16}
\end{table}

\subsection{Results of $M=4\times4$}
\label{Appendix-LS}

Here we show the results for the $M=4\times4$ case presented in Sec.~\ref{7-1}. 
Table \ref{tab:LSM16} summarizes the transfer learning results for the $M=4\times4$.
Results where pre-training is listed as N/A represent the conventional approach. Convergence is defined as $\Delta E < 1.0 \times 10^{-1}$. 

First, Table~\ref{tab:LSM16}, upper panel, summarizes the number of epochs to convergence and the initial values for transfer learning with respect to electron density $\tilde{N}$. 
These results demonstrate that high generalization performance (accuracy of the first-order perturbation approximation) is achieved even with varying electron density, indicating the effectiveness of transfer learning.

Furthermore, the bottom panel of Table~\ref{tab:LSM16} presents the results of transfer learning with respect to $\tilde{U}_{k}$. These results show that greater efficiency gains are achieved compared to the $M=2\times2$ case, suggesting that our method is more effective for larger systems. 

\subsection{Stability of Transfer Learning with MCMC}
\label{Appendix-M16-stab}

Finally, we show the results on the stability of transfer learning when using MCMC with the $M=4\times4$ Hubbard model. Compared to the $M=2\times2$ case with full bases, optimization becomes less stable when employing MCMC. Therefore, we numerically evaluate the extent to which our method is affected by this instability. 

\begin{figure}[t]
\centering
\includegraphics[width=.5\linewidth]{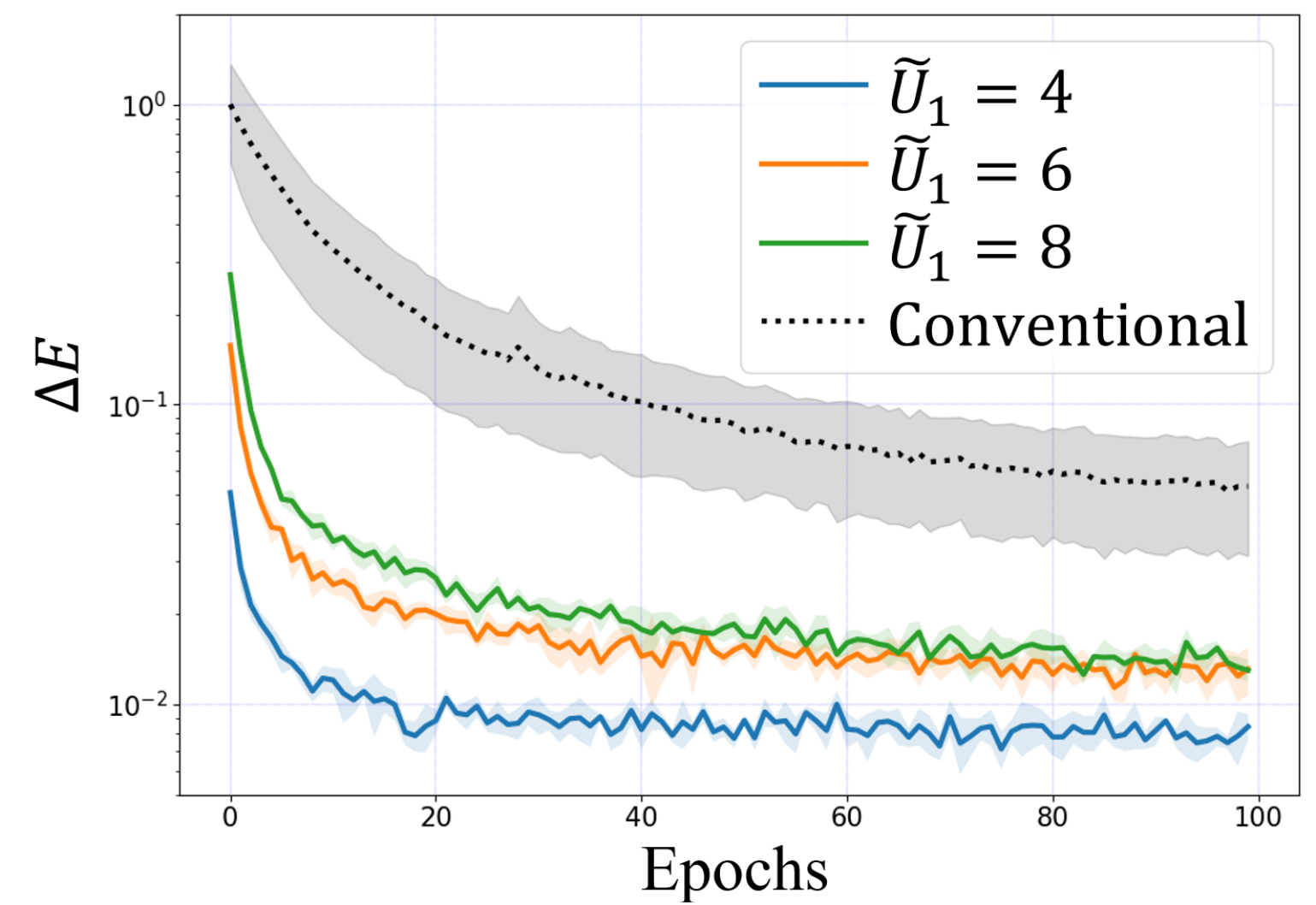}
\caption{Loss curves for targeting $\tilde{U}_{2}=2$. Results obtained using a pre-trained NQS with $\tilde{U}_{1} = 4, 6,$ and $8$ are compared to the conventional approach. The figure also shows the mean and standard deviation of $E_{\theta_{k},\varphi_{k}}$ obtained from independent five VMC calculations for each case.}
\label{fig:M16-tl-stab}
\end{figure}

Fig.~\ref{fig:M16-tl-stab} presents the stability evaluation results obtained using several pre-trained NQS $\tilde{U}_{1}=4.0,6.0,8.0$, with $\tilde{U}_{2}=2.0$ as the target parameter set. For each pre-trained NQS, five trials were conducted. These results demonstrate that optimization stability is largely maintained across almost all transfer learning instances. These findings suggest that our method exhibits high stability even in large-scale systems employing MCMC sampling.

\begin{figure}[h!]
\centering
\includegraphics[width=.5\linewidth]{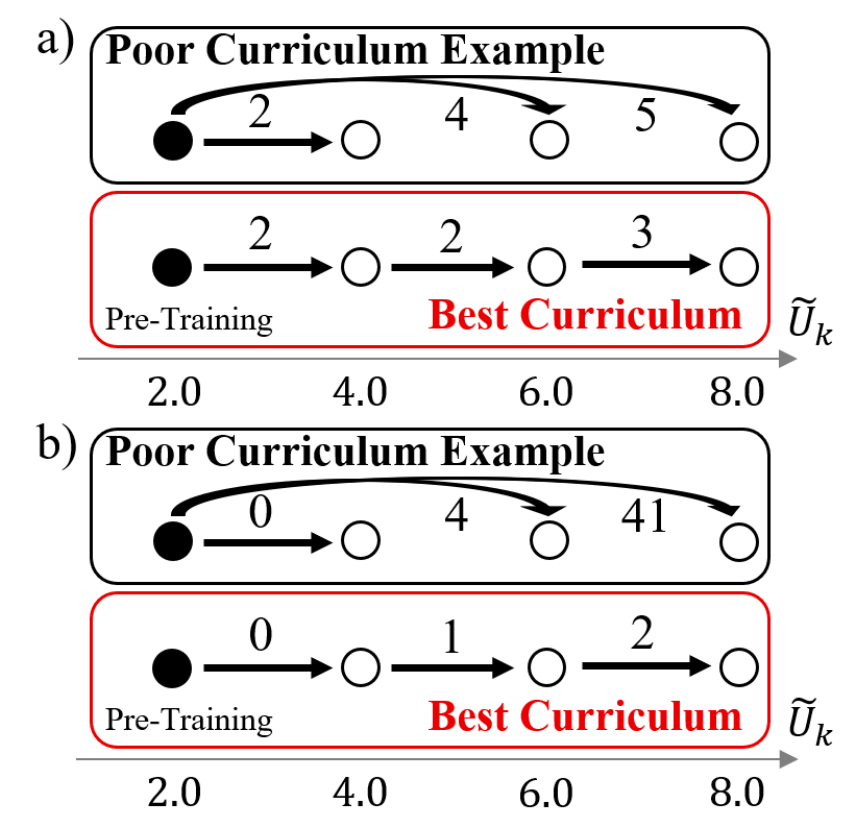}
\caption{Comparison of the best and a randomly selected poor curriculum for a) $M=2\times2$ and b) $M=4\times4$ Hubbard models. The number of epochs for each transfer learning step is indicated; arrows show the transfer direction ($\tilde{U}_1 = 2.0$ starting point).}
\label{fig:HCLresult}
\end{figure}

\begin{figure}[h!]
\centering
\includegraphics[width=.9\linewidth]{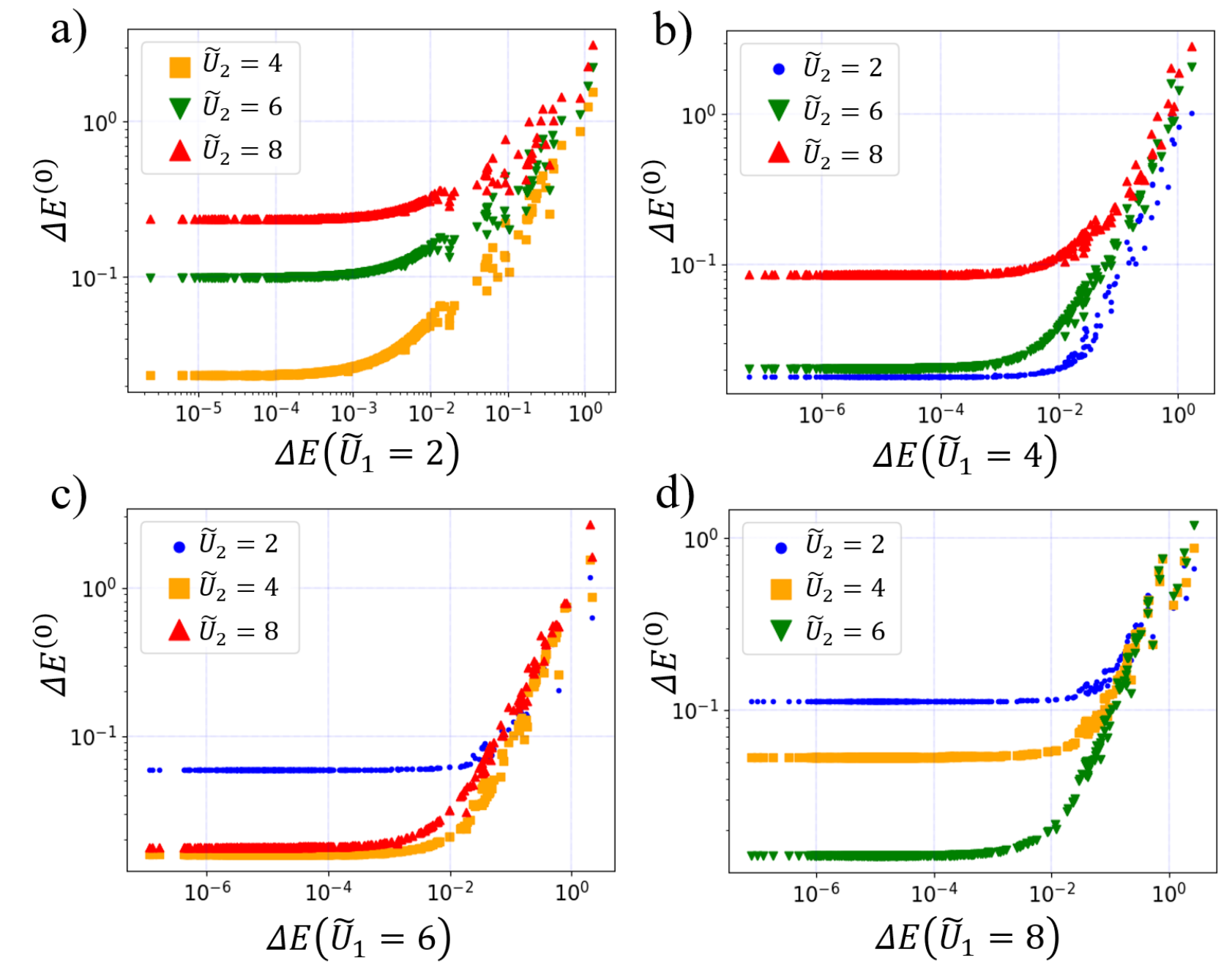}
\caption{The dependence of generalization performance, $\Delta E^{(0)}$, on the accuracy, $\Delta E(\tilde{U}_{1})$, of each pre-trained NQS. Evaluations were conducted using various pairings of pre-training $\tilde{U}_1$ values and target $\tilde{U}_2$ values within the transfer. Panels a), b), c), and d) correspond to pre-training $\tilde{U}_1$ values of $2, 4, 6$, and $8$, respectively.}
\label{fig:accpre}
\end{figure}

\subsection{Results of Explorative Curriculum Learning}
\label{Appendix-DRCL}

This section presents results from several curriculum learning schemes, leveraging the transfer learning results described above. We confirm that the curriculum proposed in Section~\ref{5-3} exhibits the highest efficiency among all curricula considered.

Fig.~\ref{fig:HCLresult} presents the results of curriculum learning for the $M=2\times2$ and $M=4\times4$ Hubbard models, showing the number of epochs required for transfer learning. These results confirm that the curriculum following the rules outlined in Section~\ref{5-3} is more efficient than any random curriculum. Furthermore, given that transfer learning with large parameter changes is less efficient in larger systems compared to smaller ones (e.g., Transfer learning from $\tilde{U}_{1}=2$ to $\tilde{U}_{2}=8$ requires 5 epochs for $M=2\times2$ and 41 epochs for $M=4\times4$, despite the looser convergence criterion for $M=4\times4$), our proposed rules are suggested to play a crucial role in enhancing efficiency for large-scale systems. We note that although $K$ is limited to 4 in this study, theoretical analysis in Section~\ref{5-3} and our results suggest that our method becomes more effective as $K$ increases.

\section{Accuracy of the Pre-Trained NQS}
\label{Appendix-pretrained}

Here, we investigate the effect of pre-trained NQS accuracy on the generalization performance of transfer learning. In Section~\ref{7-1}, the relative error of the NQS pre-trained with $\tilde{U}_{1}$ is approximately $1.0 \times 10^{-4}$ or less. However, depending on the accuracy of this $\Delta E$, the generalization performance might not be sufficiently acquired, potentially hindering the effectiveness of transfer learning. To address this, we evaluate the generalization performance, $\Delta E^{(0)}$, of pre-trained models with varying levels of accuracy ($\Delta E(\tilde{U}_{1})$), thereby determining the required accuracy of the pre-trained NQS to achieve adequate generalization.

Fig.~\ref{fig:accpre} illustrates the relationship between the accuracy of the pre-trained NQS and its generalization performance. The results indicate that, across all regimes, the generalization performance, $\Delta E^{(0)}$, plateaus at approximately $\Delta E (\tilde{U}_{1})< 1.0 \times 10^{-4}$ (this plateau represents the limitation of the accuracy of the first-order perturbation approximation when the exact solution of the ground state is employed). This suggests that sufficient generalization is achieved when this level of accuracy is ensured. Furthermore, the linear improvement in generalization performance even in the regime where $\Delta E(\tilde{U}_{1}) > 1.0 \times 10^{-4}$ indicates that a degree of generalization is attainable even when the pre-training is not fully completed. This finding suggests that transfer learning remains effective even in regimes with high system complexity and less-than-perfect accuracy (this trend is also confirmed in Section~\ref{7-3} and~\ref{Appendix-LS}).

\section{Transferable NQS with Slater Determinants}
\label{Appendix-SD}

\begin{table*}[t]
    \centering\small
    \caption{Transfer learning results from $\tilde{U}_{1}$ to $\tilde{U}_{2}$ with fixed electron density $\tilde{N}=1.0$ using SD-Net (SDN) and Pairing-Net (PN). Results where pre-training is listed as N/A represent the conventional approach. Convergence is defined as $\Delta E < 1.0 \times 10^{-4}$. The table shows the Mean and standard deviation (std.) of the number of epochs to convergence, and the convergence count (CC), based on $1000$ independent trials.}
    \vspace*{2mm}\small
    \begin{tabular}{ll|rr|rr}
        \multicolumn{1}{c}{$\tilde{U}_{1}$} & \multicolumn{1}{c}{$\tilde{U}_{2}$} & \multicolumn{1}{c}{$\mathrm{Mean}\pm\mathrm{Std.}$ (SDN)} & \multicolumn{1}{c}{$\mathrm{Mean}\pm\mathrm{Std.}$ (PN)} & \multicolumn{1}{c}{CC (SDN)} & \multicolumn{1}{c}{CC (PN)}  \\
        \midrule
        N/A & $2.0$ & $482 \pm 220$ & $61 \pm 95$ &  $400$ & $19$ \\
        $4.0$ & $2.0$ & $4.00 \pm 0.00$ & $2.00 \pm 0.00$ &  $1000$ & $1000$\\
        $6.0$ & $2.0$ & $7.00 \pm 0.00$ & $4.00 \pm 0.00$ &  $1000$ & $1000$\\
        $8.0$ & $2.0$ & $24.0 \pm 0.00$ & $5.00 \pm 0.00$ &  $1000$ & $1000$\\
        \midrule
        N/A & $4.0$ & $320 \pm 191$ & $82 \pm 26$ &  $702$ & $70$ \\
        $2.0$ & $4.0$ & $4.00 \pm 0.00$ & $2.00 \pm 0.00$ &  $1000$ & $1000$\\
        $6.0$ & $4.0$ & $4.00 \pm 0.00$ & $2.00 \pm 0.00$ &  $1000$ & $1000$\\
        $8.0$ & $4.0$ & $10.0 \pm 0.00$ & $3.00 \pm 0.00$ &  $1000$ & $1000$\\
        \midrule
        N/A & $6.0$ & $270 \pm 183$ & $83 \pm 15$ &  $776$ &  $293$\\
        $2.0$ & $6.0$ & $14.0 \pm 0.00$ & $3.00 \pm 0.00$ &  $1000$ & $1000$\\
        $4.0$ & $6.0$ & $4.00 \pm 0.00$ & $2.00 \pm 0.00$ &  $1000$ & $1000$\\
        $8.0$ & $6.0$ & $6.00 \pm 0.00$ & $2.00 \pm 0.00$ &  $1000$ & $1000$\\
        \midrule
        N/A & $8.0$ & $264 \pm 185$ & $81 \pm 13$ &  $771$ & $415$ \\
        $2.0$ & $8.0$ & $86.0 \pm 0.00$ & $5.00 \pm 0.00$ &  $1000$ & $1000$\\
        $4.0$ & $8.0$ & $47.0 \pm 0.00$ & $4.00 \pm 0.00$ &  $1000$ & $1000$\\
        $6.0$ & $8.0$ & $6.00 \pm 0.00$ & $3.00 \pm 0.00$ &  $1000$ & $1000$\\
    \end{tabular}
    \label{tab:tl-M4-SD}
\end{table*}

This section introduces SD-Net, a transferable NQS architecture distinct from the Pairing-Net presented in Section~\ref{6}. We then demonstrate the architecture-agnostic effectiveness of our method through several numerical experiments.

Similar to Pairing-Net, SD-Net is a transferable NQS architecture for the Hubbard model, designed to facilitate the transfer of electron correlation information. While Pairing-Net utilizes the NPP wave function, SD-Net employs a neural Slater determinant (NSD) wave function, which represents a Slater determinant using a neural network. The Slater determinant (SD) inherently satisfies the antisymmetry requirement for fermions and is a widely used trial wave function for electron systems. Numerous NQS architectures based on SD have been proposed~\cite{pfau2020ab,hermann2020deep,scherbela2023variational,kim2024neural,gao2024neural}, and SD-Net represents a variant of these. Furthermore, we incorporate the NCF introduced in Section~\ref{6} into the NSD wave function.

The input to the NSD wave function is identical to that of the NPP wave function; however, the output $g_{\theta_{k}}(\bm{\xi}_{i},\delta\bm{\xi}_{i})$ is interpreted as an electron orbital function. Based on this interpretation, the NSD wave function $\Psi_{\theta_{k}}^{(\mathrm{NSD})}(x)$ is defined as follows:
\begin{equation}
\Psi_{\theta_{k}}^{(\mathrm{NSD})}(x) = \mathrm{det}[G_{\theta_{k}}],
\end{equation}
where $(G_{\theta_{k}})_{ij} \equiv g_{\theta_{k}}(\bm{\xi}_{i},\delta\bm{\xi}_{ij})$, and the NSD wave function is defined as the determinant of this matrix. In SD-Net, the trial wave function is expressed as:
\begin{equation}
\Psi_{\theta_{k},\varphi_{k}}(x) =  \Psi_{\theta_{k}}^{(\mathrm{NSD})}(x)\times\exp\left[\mathcal{C}_{\varphi_{k}}(x)\right].
\end{equation}
The following section presents the results of numerical experiments using this wave function, analogous to those in Section~\ref{7-1}. The experimental setup in this section is described in~\ref{Appendix-EXset}.

Table~\ref{tab:tl-M4-SD} presents the results of transfer learning using SD-Net. These results demonstrate significant speedup and improved stability, mirroring the findings obtained with Pairing-Net in Section~\ref{7-1}. This architecture-agnostic effectiveness of our proposed method is further corroborated by these findings.

Furthermore, a comparison with the results for Pairing-Net (Table~\ref{tab:tl-M4}) reveals a slower convergence speed, suggesting that the NPP wave function possesses greater expressiveness. While SD-Net exhibits higher stability than Pairing-Net, we remark that our method dramatically improves stability, ensuring sufficient stability even in computations employing the more expressive Pairing-Net.

\section{Results of Larger-Scale Hubbard Models}
\label{Appendix-MoreLS}

\begin{figure}[t]
\centering
\includegraphics[width=.7\linewidth]{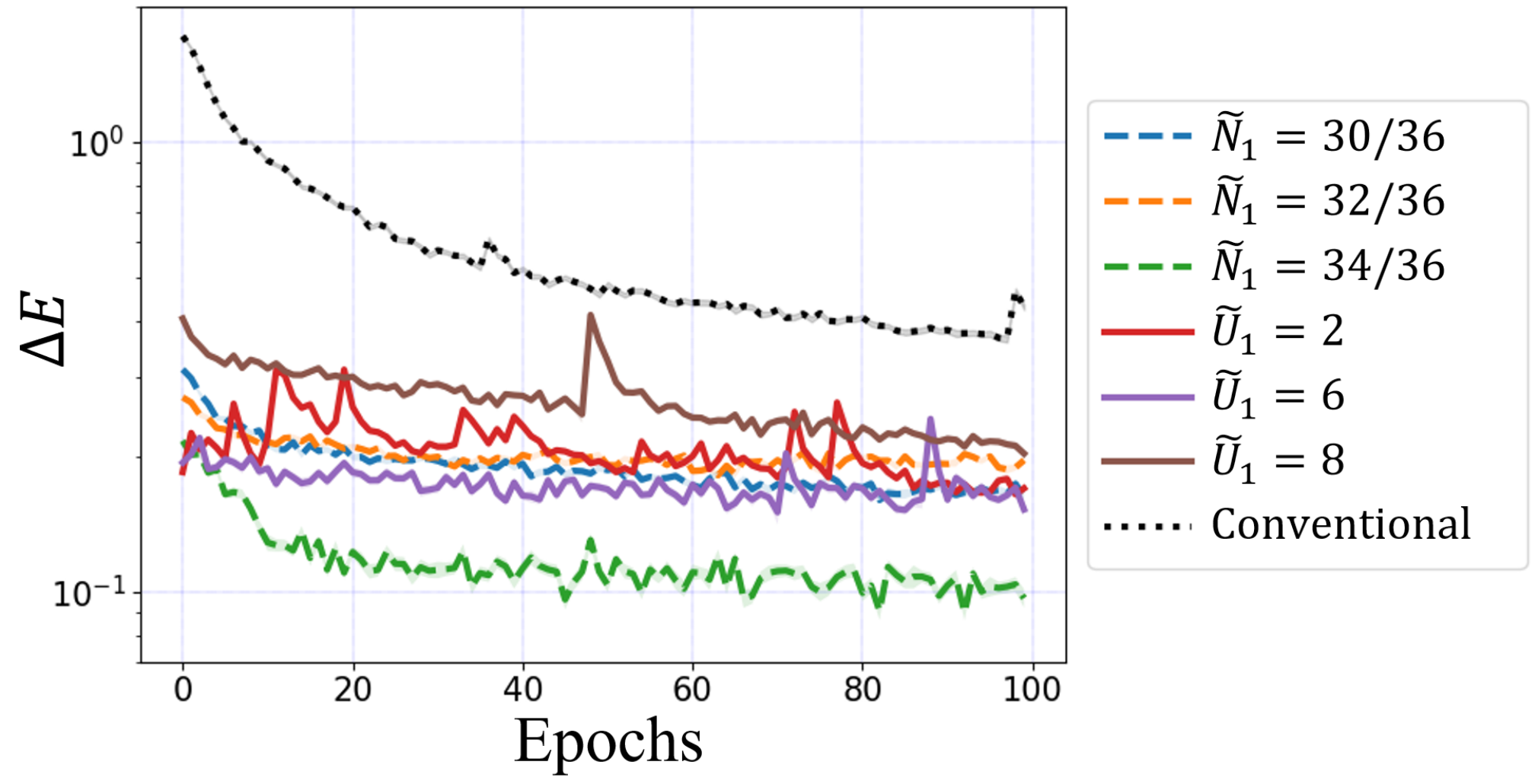}
\caption{The loss curves for $M=6\times6$ over 100 epochs applied to $(\tilde{N}_{2},\tilde{U}_{2})=(36/36,4)$ using the various pre-trained NQS.}
\label{fig:M16-NUloss-all}
\end{figure}

This section presents the results of numerical experiments conducted on a larger $M=6\times6$ Hubbard model. Similar to the $M=4\times4$ case in Section~\ref{7-1}, we investigate the effectiveness of transfer learning with Pairing-Net using several pre-trained NQS with different $\tilde{U}_{1}$ and $\tilde{N}_{1}$ values. Details of the experimental setup are provided in~\ref{Appendix-EXset}.

Fig.~\ref{fig:M16-NUloss-all} shows the loss curves for the $M=6\times6$ case, comparing transfer learning with the conventional approach.
Consistent with the findings in Section~\ref{7-1}, these results demonstrate efficiency gains from transfer learning, even in larger systems. Moreover, as predicted by the theoretical analysis in Section~\ref{5-3}, transfer to similar parameter sets shows the greatest effectiveness.
Notably, these results reveal that utilizing an electron density of $\tilde{N}_{1}=34/36$ achieves high accuracy within only a few tens of epochs. This contrasts with the findings in Section~\ref{7-1}, particularly for the $M=2 \times 2$ case, where utilizing pre-trained NQS with $\tilde{U}_{1}$ proved more effective. This discrepancy is likely attributable to finite-size effects related to electron density, as discussed in~\ref{Appendix-DRTL-Small}. Specifically, this implies that the change in electron density can be smaller in larger systems compared to smaller ones.
Moreover, despite the initial values being almost the same as $\tilde{U}_{1}=2,6$, it was confirmed that transfer learning for electron density $\tilde{N}_{1}=34/36$ converges faster. This implies that the transfer of electron density captures more complex electron correlations.
These observations suggest that our method, employing transfer learning for electron density, is more effective for large-scale systems.

\section{Results of $J_1$-$J_2$ Heisenberg Models}
\label{J1J2}

Here, we validate the applicability of our method to another challenging quantum many-body system—namely, the frustrated, strongly correlated $J_1$--$J_2$ Heisenberg model—beyond the Hubbard‐model setting. Its effective Hamiltonian reads
\begin{equation}
  \hat{H}(\bm{\lambda}) =
    J_{1}\sum_{\langle i,j \rangle}\hat{\mathbf S}_{i}\!\cdot\!\hat{\mathbf S}_{j}
  + J_{2}\sum_{\langle\langle i,j \rangle\rangle}\hat{\mathbf S}_{i}\!\cdot\!\hat{\mathbf S}_{j},
\end{equation}
where $\hat{\mathbf S}_{i}$ is the spin-$\tfrac12$ operator on site~$i$,
$\langle i,j\rangle$ ($\langle\langle i,j\rangle\rangle$) denotes a nearest-
(next-nearest-) neighbor pair, and $J_{1}>0$ ($J_{2}>0$) is the
antiferromagnetic coupling on the corresponding bond. The parameter space of the $J_1$-$J_2$ Heisenberg model can therefore be spanned by the ratio $\lambda=J_{2}/J_{1}$.

Previous studies suggest that several quantum
phase transitions occur upon varying $\lambda$; continuous transitions have been reported at $\lambda\approx0.49$ and $0.54$~\cite{PhysRevB.88.060402,PhysRevX.11.031034}. Because the change in the ground-state wave function under perturbations is inherently non-linear, it is not obvious that the curriculum rule proposed in Sec.~\ref{5-3} will remain valid. In this section, we demonstrate that our fine-tuning procedure continues to perform reliably even in the presence of these phase transitions.

\begin{figure}[t]
\centering
\includegraphics[width=.5\linewidth]{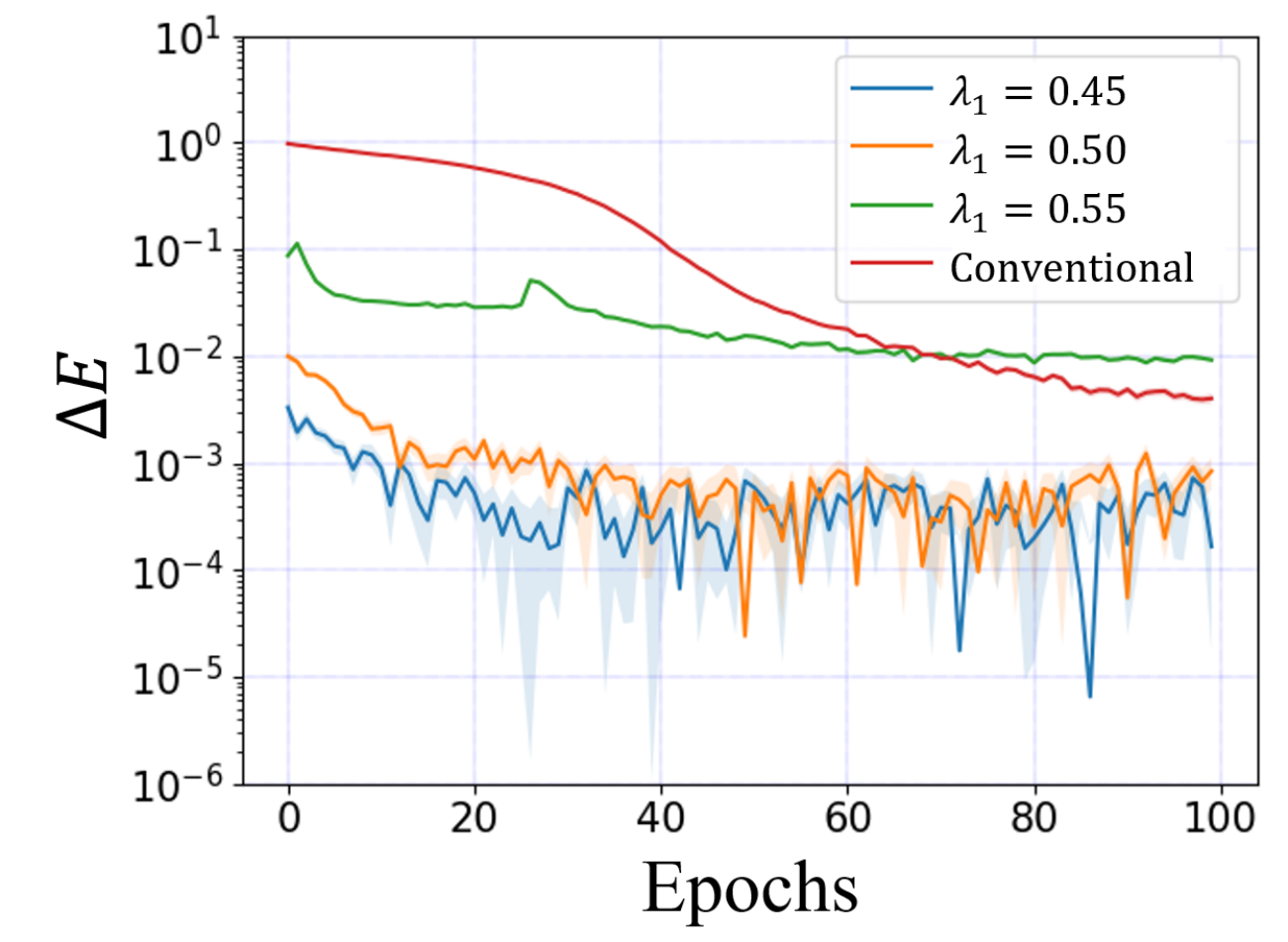}
\caption{The loss curves for $M=6\times6$ $J_1$-$J_2$ Heisenberg model over 100 epochs applied to $\lambda_{2}=0.40$ using the various pre-trained NQS.}
\label{fig:J1J2}
\end{figure}

Figure~\ref{fig:J1J2} shows the results of transfer learning using the ViT wave function~\cite{viteritti2023transformer0,rende2024simple} for the two-dimensional $J_1$-$J_2$ Heisenberg model on an $M=6\times6$ square lattice with $J_2/J_1=0.40$. Here, the ground-truth energy $E_g$ is taken from the RBM+PP results~\cite{PhysRevX.11.031034}. These results show that every fine-tuning step follows the curriculum rule proposed in Sec.~\ref{5-3} and that smaller parameter changes yield higher generalization performance during zero-shot learning than the conventional approach. In particular, transfer learning remains effective even when the optimization crosses the critical coupling ratio $J_2/J_1\approx0.49$. These findings demonstrate the broad applicability of our curriculum learning strategy to a wide class of challenging quantum many-body systems.

\newpage

\bibliographystyle{iopart-num}
\bibliography{refs}

\end{document}